\documentclass[12pt]{article}

\usepackage{amssymb,amsmath}
\usepackage{amsmath}
\usepackage{graphicx}
\usepackage{subfig}
\usepackage{multirow}
\usepackage{float}
\usepackage{cite}
\def\sech{\mathop{\rm sech}\nolimits}

\begin{document}

\begin{titlepage}

\vspace{-4cm}

\title{Heterotic Kink Solitons and their Worldvolume Action {\LARGE \\[.5cm]  }}
                       
\author{{ Burt A.~Ovrut, James Stokes} \\[5mm]
    {\it Department of Physics and Astronomy, University of Pennsylvania} \\
   {\it Philadelphia, PA 19104--6396}\\[4mm]}

\date{}

\maketitle

\begin{abstract}
\noindent
\let\thefootnote\relax\footnotetext{ovrut@elcapitan.hep.upenn.edu, ~~stokesj@sas.upenn.edu}
We present a formalism for computing the higher-order corrections to the worldvolume action of a co-dimension one kink soliton embedded in five-dimensional heterotic M-theory. The geometry of heterotic M-theory, as well as the effective theory which describes a five-brane wrapping a holomorphic curve by a topological kink in a scalar field, is reviewed. Using this formalism, the explicit worldvolume action is computed to second order in two expansion parameters--one describing the ``warp'' of the heterotic geometry and the second the fluctuation length of the soliton hypersurface. The result is expressed in terms of the trace of the extrinsic curvature and the intrinsic curvature scalar.

\vspace{.3in}
\noindent
\end{abstract}

\thispagestyle{empty}

\end{titlepage}

\section{Introduction}

There has been considerable interest in calculating the worldvolume effective action of topological solitons in various contexts. The lowest order Dirac-Born-Infeld (DBI) term was given for differing theories  in~\cite{Aganagic:1996nn,Adawi:1997sq,Sorokin:1999jx,Howe:2000vk,Derendinger:2000gy,Howe:2001wc,Cheung:2004sa,Belyaev:2010as,George:2009jn}.
In~\cite{Gregory:1990pm,Carter:1994ag}, Gregory and collaborators presented a compelling formalism for computing higher-order corrections to the action of co-dimension one scalar  ``kink'' solitons in the ``probe brane'' limit.
This involves a series expansion in a parameter $\epsilon$, the ratio of the kink thickness to the typical worldvolume fluctuation length. Using this method, the explicit worldvolume action of a probe kink in a flat background bulk space was computed~\cite{Carter:1994ag} to second order. Recently~\cite{KOS}, this formalism was modified and extended  to calculate the explicit 
higher-order terms in the worldvolume action of a scalar kink soliton in anti-deSitter (AdS) spacetime. This was carried out to second order in $\epsilon$ and a second parameter $\delta$--the ratio of the kink thickness to the radius of the AdS space. The result contains, in addition to the usual
(DBI) interaction, three higher-dimensional terms. These are proportional to $\hat{K}$, ${\hat{R}}^{(4)}$ and ${\hat{K}}^{2}$, where $\hat{K}$ and ${\hat{R}}^{(4)}$ are the intrinsic and extrinsic scalar curvatures of the worldvolume respectively. The DBI, $\hat{K}$ and ${\hat{R}}^{(4)}$ terms are the $L_{2}$, $L_{3}$ and $L_{4}$ conformal Galileons~\cite{Dvali:2000hr,Nicolis:2008in,Trodden:2011xh,Khoury:2011da,Endlich:2010zj,deRham:2010eu,Hinterbichler:2010xn,Goon:2010xh} computed with explicit coefficients. However, ${\hat{K}}^{2}$ is not a Galileon, and was shown to be of comparable magnitude in any region of temporal/spatial gradients.

The formalism developed in~\cite{KOS} allows one to explicitly compute the worldvolume effective actions of co-dimension one solitons in a much wider range of physical theories--such as superstrings and M-theory.  Of particular interest is the  five-brane soliton of M-theory; see, for example,~\cite{Duff:1994an,Stelle:1996tz}. When compactified on a small radius Calabi-Yau threefold times an $S^{1}/{\mathbb{Z}}_{2}$ interval of greater length, M-theory gives rise to heterotic M-theory~\cite{a,b,c}. This consists of a five-dimensional bulk spacetime with two boundary walls--the observable and hidden sectors respectively--with one or more co-dimension one domain walls corresponding to  five-brane solitons wrapped on holomorphic curves in the Calabi-Yau manifold~\cite{d,e,f}. Heterotic M-theory is particularly compelling, since the gauge connection in the observable sector~\cite{Donagi:1998xe,Donagi:1999gc,Buchbinder:2002ji,Buchbinder:2002pr,A,B,C,D} can be chosen so that the low energy spectrum of the theory is precisely that of the minimal supersymmetric standard model (MSSM)--that is, three families of quarks/leptons with one pair of Higgs-Higgs conjugate superfields--along with three right-handed neutrino supermultiplets, one per family~\cite{g,h,Braun:2005zv}. The topological and four-dimensional worldvolume spectrum of a single five-brane wrapped on a holomorphic curve was analyzed in~\cite{d}. However, the explicit description of this wrapped soliton in terms of the field content of heterotic M-theory has not yet been presented. 

As a first approach to this problem, an effective heterotic theory was presented in~\cite{Antunes:2002hn}. This consists of the bulk space metric and dilaton of heterotic M-theory augmented by an extra scalar field $\chi$. Topologically charged boundary walls represent the observable and hidden sectors. In addition, a specific potential energy for the dilaton and $\chi$ is added to the effective bulk Lagrangian. Using the associated BPS equations, it is shown that this theory admits a kink soliton solution which preserves four-dimensional $N=1$ supersymmetry. The solution depends only on the fifth coordinate of the bulk space and represents the M-theory five-brane wrapped on a holomorphic curve. In this paper, we will use the effective five-dimensional heterotic M-theory presented in~\cite{Antunes:2002hn} and extend the BPS kink solution to include supersymmetry breaking dependence on the four worldvolume coordinates. This is accomplished using the expansion formalism developed in~\cite{KOS}. We then explicitly compute the higher-order corrections to the worldvolume effective action to second order in the two expansions parameters.

Specifically, we will do the following. In Section 2, the five-dimensional heterotic theory presented in~\cite{Antunes:2002hn} is reviewed. In the absense of the scalar $\chi$, this is exactly the metric and dilaton sector of heterotic M-theory. Imposing an ansatz for these fields and the appropriate boundary conditions, we present the solution of the equations of motion, first found in~\cite{b,c}, that is sourced at the two boundaries and preserves $N=1$ four-dimensional supersymmetry. 
The associated BPS equations are then given and this solution is shown to satisfy them, as it must. Using the simpler BPS equations, solutions for the metric and dilaton in two different coordinate ``gauges'' are presented, as well as an analysis of the range of the associated fifth bulk space coordinate. This will be the background geometry in which the kink solution of the scalar $\chi$ will be embedded. Following~\cite{Antunes:2002hn}, the scalar field $\chi$ is introduced, along with a potential for both the dilaton and $\chi$. It is shown, using a specific gauge, that the $\chi$ equation of motion admits a topological kink soliton as a solution. It is this kink that models the wrapped heterotic five-brane.

The authors of~\cite{Antunes:2002hn} go on to solve for the metric and dilaton, including the backreaction from the kink soliton. This, however, is not what is required for our analysis. Backreaction greatly complicates the calculation of the soliton worldvolume action. Instead, following~\cite{Gregory:1990pm,Carter:1994ag} and~\cite{KOS}, we will employ the ``probe brane'' limit; that is, the kink soliton living in the pure heterotic geometry without backreaction. It is necessary to prove that such a probe limit is well-defined within the context of heterotic M-theory. This is shown in detail in the first subsection of Section 3. Having established this, we then introduce the $\epsilon$ expansion--as well as an expansion in a second parameter $\delta$--and use it to solve for non-supersymmetric kink solutions that, in addition to the fifth bulk coordinate, also depend on the four worldvolume coordinates. Following~\cite{Gregory:1990pm,Carter:1994ag,KOS}, this is accomplished within the context of Gaussian normal coordinates. The associated equations for the metric and extrinsic curvature are presented, along with the equations of motion for both the dilaton and $\chi$. The metric and extrinsic curvature equations are solved first. As discussed in~\cite{KOS}, the difficulty of solving the extrinsic curvature equation is greatly reduced by Weyl rescaling to a ``flat'' metric variable. This equation then dramatically simplifies and the new metric and extrinsic curvature are easily solved for. One then scales back to the original variables--thus solving the problem. This is accomplished to first order in $\epsilon$.

Using these results, in the final subsection of Section 3 we solve the scalar equations of motion, beginning with the dilaton. The order $\epsilon^{0}$ dilaton equation is presented and explicitly solved. We then find the exact solution for the order $\epsilon^{1}$ dilaton equation and discuss its properties. Importantly, it is demonstrated that this solution is equivalent to finding the ``off-shell'' relationship between the original fifth bulk coordinate and the Gaussian normal/worldvolume coordinates. The order $\epsilon^{0}$ and order $\epsilon^{1}$ equations for $\chi$ are then presented. The order $\epsilon^{0}$ equation is shown to admit the topological kink solution found above, as it must. The $\epsilon^{1}$ equation, however, is considerably more involved and can only be solved numerically. We do this for several canonical choices of parameters and present the results.

In Section 4, we present the formalism required to compute the four-dimensional worldvolume action of a kink hypersurface embedded in the heterotic background geometry. Inserting results from the previous sections, the action is then calculated using the $\epsilon$-expansion--associated with the fluctuation length of the worldvolume--as well as an expansion in the parameter $\delta$--measuring the ``warp'' in the heterotic geometry. We work to second order in these parameters. To make the expressions more tractable, several additional simplifying assumptions are made. First, we set two, a priori arbitrary, parameters to specific canonical values; second, we place the kink soliton at the center of the fifth dimensional interval; and third, we assume that both the 
``warp radius'', $1/\alpha$, and the worldvolume fluctuation length, $L$, lie outside this interval--that is $1/\alpha,L>\pi\rho/2$. Doing this provides a natural cut-off for all integrals at the heterotic boundary walls. Using these physically reasonable assumptions, we explicitly compute the kink hypersurface effective Lagrangian. As in the AdS case \cite{KOS}, we find that, in addition to the DBI term, there are three higher-derivative contributions. These are proportional to ${\hat{K}}$, ${\hat{R}}^{(4)}$ and ${\hat{K}}^{2}$, where ${\hat{K}}$, ${\hat{R}}^{(4)}$ are the trace of the extrinsic curvature and the intrinsic scalar curvature respectively. The coefficients of these terms, as well as the overall ``brane tension'', depend on three parameters--the width of the kink, $l$, the warp, $\delta$, and the location of the boundary walls, $u_{0}$. For a chosen $u_{0}$, graphs of the coefficients as functions of $\delta$ and $l$ are presented. Finally, we partially relax our final assumption, allowing $L$ to become smaller than $\pi\rho/2$. In this case, one must cut off all integrals at $\pm 1/\epsilon$. We find that only the coefficient of the ${\hat{K}}$ term changes substantially. A graph of this coefficient for differing values of $\epsilon$ is presented.

\section{Kink Solitons in d=5 Heterotic Spacetime}

\subsection*{Pure Heterotic Geometry}

We begin by reviewing some properties of $d=5$, $N=1$ supersymmetric heterotic M-theory. In the absence of any bulk three-branes, the bosonic action is given by
\begin{eqnarray}
&&{\cal{S}}=-\frac{1}{2\kappa_{5}^{2}} \Big( \int_{M_{5}}{d^{4}xdy\sqrt{-g}\big( \frac{1}{2}R+\frac{1}{4} g^{mn}\partial_{m}\phi \partial_{n}\phi+\frac{1}{3} \alpha^{2} e^{-2\phi} \big)}  \nonumber \\
&&\qquad \qquad +\int_{M_{4}^{(1)}} d^{4}x\sqrt{-g}2\alpha e^{-\phi}-\int_{M_{4}^{(2)}} d^{4}x\sqrt{-g}2\alpha e^{-\phi} \Big) \ ,\label{1}
\end{eqnarray}
where $g_{mn}$ is the five-dimensional metric and $\phi$ is the dimensionless dilaton. The fifth coordinate $y$ of the bulk space $M_{5}$ labels the interval between the fixed points $0$ and $\pi\rho$ of the orbifold $S^{1}/{\mathbb{Z}}_{2}$, where $\rho$ is the radius of $S^{1}$.  $M_{4}^{(1)}$ and $M_{4}^{(2)}$ are the four-dimensional orbifold planes located at $y=0$ and $y=\pi \rho$ respectively. Finally, parameters $\kappa_{5}$ and $\alpha$ are the dimension $-3/2$ Planck constant and the dimension $1$ boundary charge. 

It is straightforward to derive the order two Einstein and dilaton equations of motion in the bulk spacetime. These must be solved subject to appropriate boundary conditions. Imposing the ansatz
\begin{equation}
ds^{2}=e^{2A(y)}dx^{\mu}dx^{\nu}\eta_{\mu\nu}+e^{2B(y)}dy^{2}, \quad \phi=\phi(y) 
\label{2}
\end{equation}
it was shown in \cite{b,c} that a solution is given by
\begin{equation}
e^{2A(y)}=a_{0}^{2}h(y), \quad e^{2B(y)}=b_{0}^{2}h(y)^{4}, \quad e^{\phi(y)}=b_{0}h(y)^{3} 
\label{3}
\end{equation}
where
\begin{equation}
h(y)=-\frac{2}{3}(\alpha y+c_{0}) 
\label{4}
\end{equation}
and $a_{0},b_{0}$ and $c_{0}$ are real constants. Since $e^{2A}>0$, $h(y)$ must be positive over the entire range of $y$. Hence, $c_{0}$ must be chosen so that
\begin{equation}
c_{0}< -\alpha \pi \rho \ .
\label{5}
\end{equation}
Similarly, $e^{\phi}>0$ requires 
\begin{equation}
b_{0}>0 \ .
\label{6}
\end{equation}
Note that when computed on $S^{1}/ \mathbb{Z}_{2}$, 
\begin{equation}
\partial_{y}^{2} h=-\frac{4}{3} \alpha \big(\delta(y)-\delta(y-\pi\rho)\big) \ ,
\label{7}
\end{equation}
showing that this solution is sourced at both boundaries with the appropriate charge. Furthermore, by examining the variation of the associated fermions it was shown in \cite{b,c} that this solution preserves half of the $d=5$, $N=1$ supersymmetries, leaving a four-dimensional $N=1$ supersymmetry unbroken. Thus, heterotic M-theory arises as a BPS double domain wall vacuum of the effective action \eqref{1}.

To find solutions that preserve $d=4$, $N=1$ supersymmetry, it is far simpler to use the associated first order BPS equations. For the ansatz \eqref{2},
the BPS equations are given by
\begin{equation}
e^{-B}A^{\prime}=-\frac{\alpha}{3} e^{-\phi}, \quad e^{-B}\phi^{\prime}=-2\alpha e^{-\phi} \ . 
\label{8}
\end{equation}
The signs on the right-hand side of each equation have been chosen so as to automatically satisfy the correct boundary conditions. It is straightforward to show that any solution of \eqref{8} also satisfies the second order equations of motion, but, of course, not vice-versa.
To solve \eqref{8}, first note that for any function $B$ they imply 
\begin{equation}
A=\frac{1}{6}\phi+A_{0} \ ,
\label{9}
\end{equation}
where $A_{0}$ is a constant. To go further, however, one must specify the form of $B$. It is clear from \eqref{2} that this is a ``gauge'' choice, dependent on which coordinate $y$ is used.
As a check on this formalism, consider the canonical orbifold coordinate $y \in [0,\pi\rho]$ and take 
\begin{equation}
e^{2B(y)}=b_{0}^{2}h(y)^{4}, \quad h=-\frac{2}{3}(\alpha y+c_{0}) \ .
\label{10}
\end{equation}
It then follows from the second equation in \eqref{8} and \eqref{9} that 
\begin{equation}
e^{\phi(y)}=b_{0}h(y)^{3} , \quad e^{2A(y)}=a_{0}^{2}h(y)
\label{11}
\end{equation}
respectively, where 
\begin{equation}
a_{0}^{2}=e^{2A_{0}}b_{0}^{\frac{1}{3}} \ .
\label{12}
\end{equation}
That is, \eqref{3},\eqref{4} is a solution of the BPS equations \eqref{8}, as it must be. This is the gauge in which the geometry of heterotic $M$-theory was originally expressed \cite{b,c}. 

Another useful coordinate choice is to take
\begin{equation}
B=0 \ .
\label{13}
\end{equation}
In this gauge the metric and dilaton are of the form
\begin{equation}
ds^{2}=e^{2A(z)}dx^{\mu}dx^{\nu}\eta_{\mu\nu}+dz^{2}, \quad \phi=\phi(z)  \ .
\label{13A}
\end{equation}
It is clear from \eqref{2} and \eqref{10} that this will be the case for any coordinate $z$ defined by
\begin{equation}
dz=b_{0}h(y)^{2}dy \ .
\label{14}
\end{equation}
It is convenient to choose the coefficients so that $z$ has the same range as $y$, that is, $z\in[0, \pi\rho]$. This is easily accomplished by the appropriate choice of the integration constant and by demanding that $b_{0}$ and $c_{0}$ satisfy
\begin{equation}
(\alpha\pi\rho +c_{0})^{3}-c_{0}^{3}=\frac{27\alpha\pi\rho}{4b_{0}} \ .
\label{15}
\end{equation}
In order for $c_{0}$ to satisfy condition \eqref{5}, one must choose the negative root of the quadratic equation and restrict 
\begin{equation}
b_{0}<\frac{27}{4(\alpha\pi\rho)^{2}} \ .
\label{16}
\end{equation}
It then follows that
\begin{equation}
z=\frac{4b_{0}}{27\alpha}\big((\alpha y+c_{0})^{3}-c_{0}^{3}  \big) \ ,
\label{17}
\end{equation}
which has the desired range.

The BPS equations in $B=0$ gauge are easily solved. First, note that the second equation in \eqref{8} simply integrates to
\begin{equation}
e^{\phi(z)}=-2\alpha z+e_{0}   
\label{18}
\end{equation}
for any $z$ satisfying \eqref{14}. Specifying $z$ to be \eqref{17}, the constant is fixed by noting that 
$z=0$ implies $y=0$ and, hence,
\begin{equation}
e_{0}=e^{\phi(0)}=b_{0}h(0)^{3}=-\frac{8b_{0}c_{0}^{3}}{27} \ .
\label{19}
\end{equation}
It follows that 
\begin{equation}
e^{\phi(z)}=-2\alpha z-\frac{8b_{0}c_{0}^{3}}{27} \ .
\label{20}
\end{equation}
Note from constraint \eqref{15} that this expression is always positive, as it must be.
Inserting this into \eqref{9}, one finds
\begin{equation}
e^{2A(z)}=a_{0}^{2}b_{0}^{-1/3}\big( -2\alpha z-\frac{8b_{0}c_{0}^{3}}{27} \big)^{1/3}
\label{21}
\end{equation}
where we have used \eqref{12}. As a check, let us express $h(y)$ in \eqref{10} in terms of the coordinate $z$ given in \eqref{17}. We find that
\begin{equation}
h(y)= b_{0}^{-1/3}\big( -2\alpha z-\frac{8b_{0}c_{0}^{3}}{27} \big)^{1/3} \ .
\label{22}
\end{equation}
Hence, \eqref{20} and \eqref{21} are simply  $b_{0}h^{3}$ and $a_{0}^{2}h$ written in terms of the coordinate $z$.

It is convenient to simplify notation by defining 
\begin{equation}
{\cal{C}}=a_{0}^{2}(\frac{b_{0}}{2})^{-1/3}, \quad {\cal{D}}=-\frac{4b_{0}c_{0}^{3}}{27} \ .
\label{22A}
\end{equation}
Equations \eqref{20} and \eqref{21} then become
\begin{eqnarray}
&&e^{\phi(z)}=2(-\alpha z+{\cal{D}}) \ , \label{22B}\\
&&e^{2A(z)}={\cal{C}}(-\alpha z+{\cal{D}})^{1/3}
\label{22C}
\end{eqnarray}
respectively. The coefficients ${\cal{C}}$ and ${\cal{D}}$ are subject to contraints \eqref{6}, \eqref{16} for $b_{0}$ and \eqref{5}, \eqref{15} for $c_{0}$, but otherwise are arbitrary. Be that as it may, it is useful to note that choosing
\begin{equation}
{\cal{D}}=\frac{1}{2}~~\Rightarrow ~~\phi(z) \stackrel{\alpha \rightarrow 0}{\longrightarrow} 0
\label{22D}
\end{equation}
and
\begin{equation}
{\cal{C}}=\frac{1}{{\cal{D}}^{1/3}}~~\Rightarrow~~A(z) \stackrel{\alpha \rightarrow 0}{\longrightarrow} 0 \ .
\label{22E}
\end{equation}
It follows that the heterotic geometry will have both vanishing dilaton and a flat metric in the $\alpha \rightarrow 0$ limit if one takes
\begin{equation}
{\cal{C}}=2^{1/3}~, \quad {\cal{D}}=\frac{1}{2} \ .
\label{22F}
\end{equation}
We will use these values when doing explicit numerical computations below.

Although we will work predominantly in $B=0$ gauge, it will be useful at one point in our calculation to go to a third coordinate $z^{\prime}$ for which
\begin{equation}
B=A \ .
\label{23}
\end{equation}
This puts the metric in manifestly conformally flat form. In $B=0$ gauge, the metric can be written as
\begin{equation}
ds^{2}=e^{2A(z)}\big( dx^{\mu}dx^{\nu}\eta_{\mu\nu}+e^{-2A}dz^{2}  \big) \ ,
\label{24}
\end{equation}
where $e^{2A(z)}$ is given in \eqref{21}. It then follows that $z^{\prime}$ is defined by
\begin{equation}
dz^{\prime}=e^{-A(z)}dz \ .
\label{25}
\end{equation}
Happily, we will only need this defining relationship in the following.

\subsection*{Including Bulk Three-Branes}

As discussed in \cite{d,e}, the requirement of anomaly cancellation can necessitate introducing
$d=4$, $N=1$ preserving 3-branes into the bulk space. A fundamental description of these branes in $d=5$ heterotic spacetime has not yet been given. However, an effective theory in which a three-brane arises as a kink solution of a scalar in the heterotic geometry has been presented in \cite{Antunes:2002hn}. Let us briefly review their formalism. In addition to the metric and dilaton, the authors of \cite{Antunes:2002hn} introduced another dimensionless scalar field $\chi$. The heterotic action \eqref{1} was then generalized to
\begin{eqnarray}
&&{\cal{S}}=-\frac{1}{2\kappa_{5}^{2}} \Big( \int_{M_{5}}{d^{4}xdy\sqrt{-g}\big( \frac{1}{2}R+\frac{1}{4} g^{mn}\partial_{m}\phi \partial_{n}\phi+\frac{1}{2} e^{-\phi}g^{mn}\partial_{m}\chi \partial_{n}\chi }  \nonumber \\
&&\quad \qquad +V(\phi,\chi)\big) +\int_{M_{4}} d^{4}x\sqrt{-g}2W-\int_{M_{4}} d^{4}x\sqrt{-g}2W \Big) \ ,\label{26}
\end{eqnarray}
where 
\begin{equation}
W=e^{-\phi}\omega(\chi), \quad V(\phi,\chi)=\frac{1}{3}e^{-2\phi}\omega^{2}+\frac{1}{2}e^{-\phi}\big(\frac{d\omega}{d\chi} \big)^{2} \ .
\label{27}
\end{equation}
Note that the function $\omega$ has dimension 1. By construction, this is the bosonic part of a $d=5$, $N=1$ supersymmetric effective field theory. It is straightforward to derive the order two Einstein, dilaton and $\chi$ equations of motion in the bulk, as well as the appropriate boundary conditions. However, to find solutions that preserve $d=4$, $N=1$ supersymmetry, it is easier to use the associated first order BPS equations. For the ansatz
\begin{equation}
ds^{2}=e^{2A(y)}dx^{\mu}dx^{\nu}\eta_{\mu\nu}+e^{2B(y)}dy^{2}, \quad \phi=\phi(y), \quad \chi=\chi(y) 
\label{28}
\end{equation}
the BPS equations are given by
\begin{eqnarray}
&&e^{-B}A^{\prime}=-\frac{1}{3} e^{-\phi}\omega, \quad e^{-B}\phi^{\prime}=-2 e^{-\phi}\omega \label{29} \\
&&\qquad \qquad  e^{-B}\chi^{\prime}=\frac{\partial \omega}{\partial \chi} \ .\label{30}
\end{eqnarray}
The signs have been chosen so as to automatically satisfy the correct boundary conditions. Any solution of \eqref{29},\eqref{30} also satisfies the second order equations of motion, but not vice-versa. Note that choosing by
\begin{equation}
\omega=\alpha \ ,
\label{31}
\end{equation}
$\chi$ can be set to zero and the $A$, $\phi$ BPS conditions revert to those in the pure heterotic case \eqref{8}. 

To solve the BPS equations first note that, for any choice of  $B$ and $\omega$, equations \eqref{29} imply, once again, that 
\begin{equation}
A=\frac{1}{6}\phi+A_{0} \ ,
\label{32}
\end{equation}
where $A_{0}$ is a constant. One must now specify $\omega$. The authors of \cite{Antunes:2002hn} choose
\begin{equation}
\omega=m\chi\big(v^{2}-\frac{1}{3}\chi^{2}  \big) \ ,
\label{33}
\end{equation}
where constants $m$ and $v$ have dimensions 1 and 0 respectively. This
corresponds to introducing the potential 
\begin{equation}
V(\phi,\chi)=\frac{1}{3}e^{-2\phi}m^{2}\chi^{2}(v^{2}-\frac{1}{3}\chi^{2})^{2}+\frac{1}{2}e^{-\phi}m^{2}(v^{2}-\chi^{2})^{2}
\label{34}
\end{equation}
into the action \eqref{26}. Choosing the coordinate gauge 
\begin{equation}
B=0 \ ,
\label{35}
\end{equation}
they show that the $\chi$ equation \eqref{30} has a kink solution
\begin{equation}
{\chi}_{(0)}=v~tanh\big(mv(z-z_{0}) \big) 
\label{36}
\end{equation}
describing a static domain wall located at $z=z_{0}$ of width
\begin{equation}
l=\frac{1}{mv} \ .
\label{37}
\end{equation}
They also present the solution for the dilaton and $A$ in this gauge. These are sourced by $\omega$ in \eqref{33} and, hence, include the ``backreaction'' of the geometry to the kink. However, for this reason these exact results are not of interest here and we won't present them. Note that the kink solution \eqref{36} exists in any other gauge as well, with $z$ expressed in terms of the new coordinate. 

\section{The $\epsilon$-Expansion and Non-BPS Kinks}

In this paper, we want to generalize this static kink solution to include fluctuations of length scale $L$ in the the remaining four spacetime coordinates. As described in \cite{Gregory:1990pm,Carter:1994ag,KOS}, this will be accomplished using an expansion in the small parameter
\begin{equation}
\epsilon=\frac{l}{L} \ .
\label{38}
\end{equation}
To include the backreaction of the kink, it is necessary for consistency to generalize both the metric $g_{mn}$ and dilaton $\phi$ solutions, in addition to $\chi$. Although possible, this greatly complicates the analysis. To avoid this, in \cite{KOS} we employed the ``probe'' limit in both the flat spacetime and AdS cases, and want to use this in the heterotic theory as well. However, one must first demonstrate that the probe limit is well-defined in this context. 

\subsection*{The Probe Brane Limit}

Recall that taking $\omega=\alpha$ brings one back to the pure heterotic case without bulk three-branes. The $\phi$ and $A$ geometry is then completely specified; for example,  by \eqref{20} and \eqref{21} in the $B=0$ gauge. Now, generalize $\omega$ to 
\begin{equation}
\omega= \alpha+m\chi\big(v^{2}-\frac{1}{3}\chi^{2}  \big) \ .
\label{39}
\end{equation}
This allows the kink solution of the $\chi$ equation \eqref{30} but, since it appears on the right-hand side of each equation in \eqref{29}, introduces potential backreaction on the heterotic geometry. To analyze the probe limit, let us
re-express the BPS equations \eqref{29} and \eqref{30} in terms of dimension $\frac{3}{2}$ fields $\Phi$ and ${\tilde{\chi}}$. We find that
\begin{eqnarray}
&&e^{-B}A^{\prime}=-\frac{1}{3} e^{-\Phi/M_{P}^{3/2}}\frac{\omega}{M_{P}^{3}}, \quad e^{-B}\Phi^{\prime}=-2 e^{-\Phi/M_{P}^{3/2}}\frac{\omega}{M_{P}^{3/2}} \label{40} \\
&&\qquad \qquad \qquad  \qquad \quad e^{-B}{\tilde{\chi}}^{\prime}=\frac{\partial \omega}{\partial{\tilde{ \chi}}} \ ,\label{41}
\end{eqnarray}
where $\omega({\tilde{\chi}})$ now has dimension 4 and $M_{P}$ is the dimension 1 Planck mass defined as 
\begin{equation}
M_{P}=\kappa_{5}^{2/3} \ .
\label{42}
\end{equation}
The specific $\omega$ in \eqref{39} now becomes 
\begin{equation}
\omega= M_{P}^{3}\alpha+{\tilde{m}}{\tilde{\chi}}\big({\tilde{v}}^{2}-\frac{1}{3}{\tilde{\chi}}^{2}
 \big) \ ,
\label{43}
\end{equation}
where ${\tilde{m}}$ and ${\tilde{v}}$ have dimensions $-\frac{1}{2}$ and $\frac{3}{2}$ respectively.
Rescaling to dimensionless variables as
\begin{equation}
{\Phi} \rightarrow M_{P}^{3/2}\phi, \quad {\tilde{\chi}}\rightarrow \eta \chi, \quad {\tilde{m}}\rightarrow \frac{m}{\eta}, \quad {\tilde{v}}\rightarrow \eta v
\label{44}
\end{equation}
where $\eta$ is a dimension $\frac{3}{2}$ parameter, the BPS equations become
\begin{eqnarray}
&&e^{-B}A^{\prime}=-\frac{\alpha}{3} e^{-\phi}\big(1+(\frac{\eta^{2}}{M_{P}^{3}})  (\frac{m}{\alpha})\chi (v^{2}-\frac{1}{3}\chi^{2})\big),\nonumber \\
&&e^{-B}\phi^{\prime}=-2\alpha e^{-\phi}\big(1+(\frac{\eta^{2}}{M_{P}^{3}})  (\frac{m}{\alpha})\chi (v^{2}-\frac{1}{3}\chi^{2})\big)  \label{45} 
\end{eqnarray}
and
\begin{equation}
e^{-B}{\chi}^{\prime}=m\big(v^{2}-\chi^{2} \big) \ .
 \label{46}
 \end{equation}
Under the assumption that parameters $\eta$ and $m$ are independent of the d=5 Planck mass,
it follows that in the limit $M_{P} \rightarrow \infty$ the BPS equations simplify to
\begin{eqnarray}
&&e^{-B}A^{\prime}=-\frac{\alpha}{3} e^{-\phi}, \quad e^{-B}\phi^{\prime}=-2\alpha e^{-\phi}  \label{47} \\
&&\qquad \qquad e^{-B}{\chi}^{\prime}=m\big(v^{2}-\chi^{2} \big) \ .\label{48}
\end{eqnarray}
Note that \eqref{47} are precisely the equations for the pure heterotic geometry without three-branes, whereas \eqref{48} is the same as the $\chi$ equation which admits the kink solution. We conclude that a well-defined probe limit exists in which the kink lives in the pure heterotic background {\it without backreaction}.

Generalizing the purely $z$ dependent kink solution to include spatial fluctuations, necessarily breaks $N=1$ supersymmetry. It follows that this analysis must be conducted using the order two equations of motion. It is tedious, but straightforward, to extend the previous argument to show that the stress-energy of the $\chi$ field decouples from the order two Einstein and dilaton equations when $M_{P}\rightarrow \infty$. In this probe limit, the relevant Lagrangian for $\chi$ is given by
\begin{equation}
{\cal{L}}_{\chi}=-\frac{1}{2}\big(\frac{\eta^{2}}{2}  \big)e^{-\phi}g^{mn}\partial_{m} \chi\partial_{n}\chi -\frac{1}{2}V(\phi,\chi) \ ,
\label{49}
\end{equation}
where $g_{mn}$ and $\phi$ are fixed in the heterotic background specified by the solutions to \eqref{47}. Noting that in rescaled variables \eqref{43} becomes
\begin{equation}
\omega= M_{P}^{3} \alpha+\eta^{2} m \chi \big(v^{2}-\frac{1}{3}\chi^{2} \big) \ ,
\label{50}
\end{equation}
the potential leading to the kink solution is given by
\begin{eqnarray}
&&V(\phi,\chi)=M_{P}^{3}\frac{1}{3}\alpha^{2}e^{-2\phi}\big(1+(\frac{\eta^{2}}{M_{P}^{3}})  (\frac{m}{\alpha})\chi (v^{2}-\frac{1}{3}\chi^{2}) \big)^{2} \nonumber \\
&&\qquad \qquad+\frac{1}{2}e^{-\phi}\eta^{2}m^{2}(v^{2}-\chi^{2})^{2}-M_{P}^{3}\frac{1}{3}\alpha^{2}e^{-2\phi} \nonumber  \\
&& \stackrel{M_{P}\rightarrow \infty}{\longrightarrow} \frac{2}{3}\eta^{2} e^{-2\phi}\alpha m\chi(v^{2}-\frac{1}{3}\chi^{2})+\frac{1}{2}\eta^{2}e^{-\phi}m^{2}(v^{2}-\chi^{2})^{2} \ .
\label{51}
\end{eqnarray}
Note that we have subtracted off the part of the potential energy that acts as the source of the pure dilaton background solution.
The associated $\chi$ equation of motion is 
\begin{equation}
\Box \chi -g^{mn}\partial_{m} \phi \partial_{n} \chi-\frac{1}{\eta^{2}}e^{\phi}\frac{\partial V}{\partial \chi}=0 \ .
\label{52}
\end{equation}
Since the solutions to the BPS conditions also satisfy the second order equations of motion, it follows that for background fields $g_{mn}$ and $\phi$ of the pure heterotic geometry, equation \eqref{52} continues to admit the kink solution. In $B=0$ gauge, this is given in \eqref{36}. 

\subsection*{Gaussian Normal Coordinates}

Using the probe limit, one can now generalize this to kink solutions that depend on the remaining four spacetime coordinates as well as the fifth dimension. As in the flat spacetime  and AdS cases discussed in \cite{KOS}, this will be achieved using a perturbative expansion in the small parameter $\epsilon$ in \eqref{38}. It is most easily carried out in Gaussian normal coordinates, defined as follows. Let $\Sigma$ be the kink defect worldsheet and $n^{m}$ the unit geodesic normal to $\Sigma$.  Generalize $z-z_{0}$, where $z_{0}$ is the location of the kink in the $\epsilon \rightarrow 0$ limit, to be the proper length along the integral curves of $n^{m}$ and denote $z_{g}=z-z_{0}$. Note that this vanishes on the specific kink hypersurface. The remaining four worldsheet coordinates of $\Sigma$ will be denoted by $\sigma^{\mu}$, $\mu=0,\dots,3$.

Each constant $z_{g}$ surface has a unit normal $n_{m}$, with an intrinsic metric $h_{mn}$ and extrinsic curvature $K_{mn}$ defined as
\begin{equation}
h_{mn}=g_{mn}-n_{m}n_{n}, \quad K=h^{p}_{m}\nabla_{p}n_{n} 
\label{53}
\end{equation}
respectively.
These two quantities are not independent. The metric and extrinsic curvature are related by
\begin{equation}
{\cal{L}}_{n}h_{mn}=2K_{mn} \ ,
\label{54} 
\end{equation}
where ${\cal{L}}_{n}$ is the Lie derivative along the $n^{m}$ vector field. Furthermore, in a general curved five-dimensional spacetime there is a constraint equation on the extrinsic curvature given by
\begin{equation}
{\cal{L}}_{n}K_{mn}=K_{mp}K^{p}_{n}-R^{(5)}_{rspq}n^{s}n^{q}h^{r}_{m}h^{q}_{n} \ .
\label{55}
\end{equation}
For the heterotic background geometry defined above, we find that
\begin{equation}
R^{(5)}_{rspq}n^{s}n^{q}h^{r}_{m}h^{p}_{n}= \frac{5}{36} \frac{\alpha^{2}}{(-\alpha (z_{g}+z_{0})+{\cal{D}})^{2}}h_{mn} \ .
\label{58}
\end{equation}

Written in Gaussian normal coordinates the $\chi$ equation of motion \eqref{52} becomes
\begin{equation}
{\cal{L}}_{n}^{2}\chi+K{\cal{L}}_{n}\chi+D^{m}D_{m}\chi-{\cal{L}}_{n}\phi{\cal{L}}_{n}\chi-h^{mn}\partial_{m}\phi\partial_{n}\chi-\frac{1}{\eta^{2}}e^{\phi}\frac{\partial V}{\partial \chi}=0 \ ,
\label{59}
\end{equation}
where 
\begin{equation}
K=h^{mn}K_{mn}, \quad D_{m}=h^{p}_{m}\nabla_{p}
\label{60}
\end{equation}
and $\phi$ is the background dilaton solution given by \eqref{20} in $B=0$ gauge. The fact that the background dilaton solution enters the equation of motion of the kink scalar is new to heterotic M-theory,  and requires us to generalize the formalism of \cite{KOS}. In the probe limit where the background decouples from the $\chi$ field, the equation of motion of the dilaton $\phi$ is given by
\begin{equation}
\Box \phi+\frac{4}{3}\alpha^{2}e^{-2\phi}=0 \ .
\label{61}
\end{equation}
In terms of Gaussian normal coordinates this becomes
\begin{equation}
{\cal{L}}_{n}^{2}\phi+K{\cal{L}}_{n}\phi+D^{m}D_{m}\phi+\frac{4}{3}\alpha^{2}e^{-2\phi}=0 \ .
\label{62}
\end{equation}
This equation will allow us to express the fixed dilaton solution \eqref{20} in $B=0$ gauge in terms of Gaussian normal coordinates. 

\subsection*{The $\epsilon$-Expansion}

Scaling to dimensionless variables
\begin{equation}
u=\frac{z_{g}}{l},\quad u_{0}=\frac{z_{0}}{l}, \quad {K}_{mn}=\frac{1}{L}{\kappa}_{mn}
\label{63}
\end{equation}
 equations \eqref{54},\eqref{55},\eqref{59} and \eqref{62} become
\begin{eqnarray}
&&{h}_{mn}^{\prime}=2\epsilon {\kappa}_{mn} \ , \label{64} \\
&&\epsilon {\kappa}_{mn}^{\prime}=\epsilon^{2} {\kappa}_{mp}{\kappa}^{p}_{n} -\frac{5}{36} \frac{\delta^{2}}{(-\delta (u+u_{0})+{\cal{D}})^{2}}h_{mn}\label{65} \\
&&\chi^{\prime\prime}+\epsilon{ \kappa} \chi^{\prime}+\epsilon^{2}D^{m}D_{m}\chi-\chi^{\prime}\phi^{\prime}+\epsilon^{2}D^{m}\chi D_{m}\phi \nonumber \\
&& -2 (1-\chi^{2})(\frac{\delta}{3}-\chi e^{\phi})e^{-\phi}=0 \label{66} \\
&&\phi^{\prime\prime}+\epsilon{ \kappa} \phi^{\prime}+\epsilon^{2}D^{m}D_{m}\phi+\frac{4}{3}\delta^{2}e^{-2\phi}=0 \label{67}
\end{eqnarray}
where $^{\prime}=\frac{\partial}{\partial u}$, we have defined 
\begin{equation}
\delta=\alpha l \ .
\label{68}
\end{equation}
Note that, without loss of generality, we have set parameter $v=1$ in the potential energy term of \eqref{66}.

These equations can now be solved by expanding each dimensionless quantity as a power series in $\epsilon$. That is, let
\begin{eqnarray}
&& \chi=\chi_{(0)}+\epsilon \chi_{(1)}+\frac{\epsilon^{2}}{2}\chi_{(2)}+{\cal{O}}(\epsilon^{3}), \label{69} \\
&& \phi=\phi_{(0)}+\epsilon \phi_{(1)}+\frac{\epsilon^{2}}{2}\phi_{(2)}+{\cal{O}}(\epsilon^{3}), \label{70} \\
&& h_{mn}=h_{(0)mn}+\epsilon h_{(1)mn}+\frac{\epsilon^{2}}{2}h_{(2)mn}+{\cal{O}}(\epsilon^{3}), \label{71} \\
&& \kappa_{mn}= \frac{1}{\epsilon}\kappa_{(0)mn}+ \kappa_{(1)mn}+\frac{\epsilon}{2} \kappa_{(2)mn}+\frac{\epsilon^{2}}{6} \kappa_{(3)mn}+ {\cal{O}}(\epsilon^{3}) \label{72}
\end{eqnarray}
where each coefficient is either purely $u$-dependent or a function of all five coordinates $(\sigma^{\mu}, u)$. Substituting these into \eqref{64}-\eqref{67}, one obtains equations for each coefficient function order by order in $\epsilon$.

\subsubsection*{The Metric and Extrinsic Curvature:}

We begin our analysis by considering the $h_{mn}$ and $\kappa_{mn}$ equations, \eqref{64} and \eqref{65} respectively. Note that the $h_{mn}$ term on the right side of \eqref{65}, which arises from the non-vanishing curvature tensor \eqref{58}, indicates that the lowest order extrinsic curvature $\kappa_{(0)mn}$ must be non-vanishing. The same situation occurs in the AdS case \cite{KOS}, and greatly complicates the solution of these two coupled equations. As discussed in \cite{KOS}, this problem can be circumvented by writing these equations in terms of the rescaled metric
\begin{equation}
\tilde{h}_{mn}=e^{-2A(u)}h_{mn} \ ,
\label{73}
\end{equation}
where
\begin{equation}
e^{2A(u)}={\cal{C}}(-\delta( u+u_{0})+{\cal{D}})^{1/3}  \ .
\label{74}
\end{equation}
First consider the $h_{mn}$ equation \eqref{64}. In terms of $\tilde{h}_{mn}$, this is
\begin{equation}
\partial_{u}(e^{2A(u)}\tilde{h}_{mn})=2\epsilon\kappa_{mn} \ .
\label{75}
\end{equation}
Using \eqref{74}, it follows that
\begin{equation}
e^{2A(u)} \partial_{u} \tilde{h}_{mn}=2\epsilon \Big( \kappa_{mn}-\frac{\delta}{\epsilon}
(\frac{-{\cal{C}}^{3}}{6})(e^{-2A})^{3}h_{mn}  \Big) \ .
\label{76}
\end{equation}
As discussed in \cite{KOS}, the metric $\tilde{h}_{mn}$ is associated with the $B=A$ gauge and the coordinate $z^{\prime}$ defined in \eqref{25}. Then
\begin{equation}
\partial_{u}=(e^{-2A(u)})^{1/2}\partial_{u^{\prime}} 
\label{77}
\end{equation}
and \eqref{76} becomes
\begin{equation}
\tilde{h}^{\prime}_{mn}=2\epsilon \tilde{\kappa}_{mn} \ ,
\label{78}
\end{equation}
where
\begin{equation}
\tilde{\kappa}_{mn}= \Big( \kappa_{mn}-\frac{\delta}{\epsilon}
(\frac{-{\cal{C}}^{3}}{6})(e^{-2A})^{3}h_{mn}  \Big)(e^{-2A})^{1/2} 
\label{79}
\end{equation}
and $^{\prime}=\frac{\partial}{\partial u^{\prime}}$. Equation \eqref{78} replaces \eqref{64} and \eqref{79} defines the exact relationship between $\tilde{\kappa}_{mn}$ and $\kappa_{mn}$.

Now consider the $\kappa_{mn}$ equation \eqref{65}. Inverting expression \eqref{79} gives
\begin{equation}
\kappa_{mn}=\tilde{\kappa}_{mn}(e^{2A})^{1/2}+\frac{\delta}{\epsilon}
(\frac{-{\cal{C}}^{3}}{6})(e^{-2A})^{3}h_{mn} \ .
\label{80}
\end{equation}
Inserting this into \eqref{65}, we find that the curvature term proportional to $h_{mn}$ exactly cancels, as do several unrelated cross terms. The result is
\begin{equation}
\epsilon \partial_{u}\big(\tilde{\kappa}_{mn}(e^{2A})^{1/2}\big)=\epsilon^{2} \tilde{\kappa}_{mp}\tilde{\kappa}_{qn}\tilde{h}^{pq} \ .
\label{81}
\end{equation}
Finally, using \eqref{74} and \eqref{77} one finds
\begin{equation}
\epsilon \tilde{\kappa}_{mn}^{\prime}=\epsilon^{2}\tilde{\kappa}_{mp}\tilde{\kappa}_{qn}\tilde{h}^{pq}+\delta\epsilon (\frac{{\cal{C}}^{2}}{6})(e^{-2A})^{5/2}\tilde{\kappa}_{mn} 
\label{82}
\end{equation}
where $^{\prime}=\frac{\partial}{\partial u^{\prime}}$.
This equation replaces \eqref{65}. It is important to note that the inhomogenous term proportional to $h_{mn}$ on the right side of \eqref{65} has now been replaced by the term proportional to $\tilde{\kappa}_{mn}$. Hence, $\tilde{\kappa}_{(0)mn}$ can vanish, simplifying the remaining calculation.

We now solve the equations \eqref{78} and \eqref{82} order by order in the $\epsilon$ expansion.
First consider the $\tilde{h}_{mn}$ equation \eqref{78}. Substituting \eqref{71} and \eqref{72}  written in terms of $\sim$ variables into \eqref{78}, we find to order $\epsilon^{0}$ and $\epsilon^{1}$ that
\begin{eqnarray}
&&\tilde{h}^{\prime}_{(0)mn}=2 \tilde{\kappa}_{(0)mn} \label{83} \ , \\
&&\tilde{h}^{\prime}_{(1)mn}=2 \tilde{\kappa}_{(1)mn} \label{84} 
\end{eqnarray}
respectively. Similarly, inserting \eqref{71} and \eqref{72} into the $\tilde{\kappa}_{mn}$ equation \eqref{82}, we find to order  $\epsilon^{0}$ and $\epsilon^{1}$ that
\begin{eqnarray}
&& \tilde{\kappa}_{(0)mn}^{\prime}=\tilde{\kappa}_{(0)mp}\tilde{\kappa}_{(0)qn}\tilde{h}_{(0)}^{pq}+\delta (\frac{{\cal{C}}^{2}}{6})(e^{-2A})^{5/2}\tilde{\kappa}_{(0)mn} \label{85} \ ,\\
&& \tilde{\kappa}_{(1)mn}^{\prime}=\tilde{\kappa}_{(0)mp}\tilde{\kappa}_{(1)qn}\tilde{h}_{(0)}^{pq}+\tilde{\kappa}_{(1)mp}\tilde{\kappa}_{(0)qn}\tilde{h}_{(0)}^{pq} \label{86} \\
&&\qquad \quad+\tilde{\kappa}_{(0)mp}\tilde{\kappa}_{(0)qn}\tilde{h}_{(1)}^{pq}+\delta (\frac{{\cal{C}}^{2}}{6})(e^{-2A})^{5/2}\tilde{\kappa}_{(1)mn} \nonumber \ .
\end{eqnarray}
\\
\noindent {\it Order $\epsilon^{0}$}: Since at this order $\tilde{n}_{m}=\tilde{n}_{5}$, it follows from \eqref{73} that
\begin{equation}
\tilde{h}_{(0)mn}=\hat{\tilde{h}}_{(0)mn}(\sigma)
\label{87}
\end{equation}
with $\hat{\tilde{h}}_{(0)mn}$ unspecified. Hence, the $\tilde{h}_{(0)mn}$ equation \eqref{83} gives
\begin{equation}
\tilde{\kappa}_{(0)mn}=0 \ ,
\label{88}
\end{equation}
as expected. Note that the $\tilde{\kappa}_{(0)mn}$ equation \eqref{85} is now trivially satisfied.\\

\noindent {\it Order $\epsilon^{1}$}: First consider the $\tilde{\kappa}_{(1)mn}$ equation \eqref{86}. Instead of solving this directly, it is simpler to go back to \eqref{81}. Inserting the 
$\epsilon$-expansion using $\tilde{\kappa}_{(0)mn}=0$, it follows that
\begin{equation}
\partial_{u}(\tilde{\kappa}_{(1)mn}(e^{2A})^{1/2})=0 \ .
\label{89}
\end{equation}
Therefore,
\begin{equation}
\tilde{\kappa}_{(1)mn}=(e^{-2A})^{1/2}\hat{\tilde{\kappa}}_{(1)mn}(\sigma)
\label{90}
\end{equation}
with $\hat{\tilde{\kappa}}_{(1)mn}$ unspecified. Finally, insert \eqref{90} into the $\tilde{h}_{(1)mn}$ equation \eqref{84}. Using \eqref{74},\eqref{77} and choosing the integration constant so that  
$\tilde{h}_{(1)mn}$ has a smooth $\delta \rightarrow 0$ limit, we find that
\begin{equation}
\tilde{h}_{(1)mn}=-\frac{3}{\delta{\cal{C}}^{3}}\Big((e^{2A})^{2}-{\cal{C}}^{2}{\cal{D}}^{2/3} \Big)\hat{\tilde{\kappa}}_{(1)mn}(\sigma) \ .
\label{91}
\end{equation}

Having found $\tilde{h}_{mn}$ and $\tilde{\kappa}_{mn}$ to first order in $\epsilon$, one can now immediately determine $h_{mn}$ and $\kappa_{mn}$ to the same order. Inserting the $\epsilon$-expansions of $h_{mn}$ and $\tilde{h}_{mn}$ into \eqref{73}, and using \eqref{87} and \eqref{91}, we finds that
\begin{eqnarray}
&&h_{(0)mn}=e^{2A}\hat{\tilde{h}}_{(0)mn}(\sigma), \label{92} \\
&&h_{(1)mn}=-\frac{3}{\delta{\cal{C}}^{3}}\Big((e^{2A})^{2}-{\cal{C}}^{2}{\cal{D}}^{2/3} \Big)(e^{2A})\hat{\tilde{\kappa}}_{(1)mn}(\sigma)\  .
\label{93}
\end{eqnarray}
Similarly, $\epsilon$ expanding $\kappa_{mn}$, $\tilde{\kappa}_{mn}$ and $h_{mn}$ in the relation \eqref{80}, and using \eqref{87},\eqref{88},\eqref{90} and \eqref{91}, gives
\begin{eqnarray}
&&\kappa_{(0)mn}=-\frac{\delta{\cal{C}}^{3}}{6}(e^{-2A})^{2}\hat{\tilde{h}}_{(0)mn}(\sigma), \label{94} \\
&& \kappa_{(1)mn}=\frac{1}{2}\Big(3-{\cal{C}}^{2}{\cal{D}}^{2/3} (e^{-2A})^{2} \Big)\hat{\tilde{\kappa}}_{(1)mn}(\sigma) \ .
\label{95}
\end{eqnarray}
Finally, we must determine the trace of the extrinsic curvature
\begin{equation}
\kappa=h^{mn}\kappa_{mn} 
\label{96}
\end{equation}
order by order in the $\epsilon$-expansion. The $\epsilon^{0}$ and $\epsilon^{1}$ terms are given by
\begin{equation}
\kappa_{(0)}=h_{(0)}^{mn}\kappa_{(0)mn}~, \quad \kappa_{(1)}=h_{(0)}^{mn}\kappa_{(1)mn}+h_{(1)}^{mn}\kappa_{(0)mn}
\label{97}
\end{equation}
respectively. It follows immediately from \eqref{92} and \eqref{94} that
\begin{equation}
\kappa_{(0)}=-\frac{2\delta{\cal{C}}^{3}}{3}(e^{-2A})^{3} \ ,
\label{98}
\end{equation}
where we have used the fact that $h_{(0)}^{mn}h_{(0)mn}=4$. Similarly, using \eqref{92}-\eqref{95} and the relation $h_{(0)}^{mp}h_{(1)pn}+h_{(1)}^{mp}h_{(0)pn}=0$ we find 
\begin{equation}
\kappa_{(1)}=(e^{-2A}){\hat{\kappa}}_{(1)}(\sigma) \ ,
\label{99}
\end{equation}
with ${\hat{\kappa}}_{(1)}(\sigma)=\hat{\tilde{h}}_{(0)}^{mn}(\sigma)\hat{\tilde{\kappa}}_{(1)mn}(\sigma) $.

\subsubsection*{The Scalar Equations of Motion:}

\noindent {\it The $\phi$ Equation}-- ~In the probe limit,  the dilaton scalar $\phi$ is the fixed background field given by  \eqref{22B} in $B=0$ gauge as
\begin{equation}
e^{\phi(z)}=2(-\alpha z+{\cal{D}}) \ .
\label{100}
\end{equation}
Although $\phi$ is fixed, the fact that it enters the $\chi$ equation of motion requires one to do an $\epsilon$-expansion of the $\phi$ field as well. As we will see, the meaning of this is to fix the relationship between the $z$ coordinate and the Gaussian normal coordinate $z_{g}$ order by order in $\epsilon$. To do this, recall that the $\phi$ equation of motion in Gaussian normal coordinates is given by \eqref{67}. To order $\epsilon^{0}$ and $\epsilon^{1}$ in the $\epsilon$-expansion this becomes
\begin{eqnarray}
&&\phi_{(0)}^{\prime \prime}+\kappa_{(0)}\phi_{(0)}^{\prime}+\frac{4\delta^{2}}{3}e^{-2\phi_{(0)}}=0 \ , \label{101} \\
&&\phi_{(1)}^{\prime \prime}+\kappa_{(0)}\phi_{(1)}^{\prime}+\kappa_{(1)}\phi_{(0)}^{\prime}-\frac{8\delta^{2}}{3}e^{-2\phi_{(0)}}\phi_{(1)}=0 \label{102} 
\end{eqnarray}
respectively, where $\kappa_{(0)}$ and $\kappa_{(1)}$ are given in \eqref{98} and \eqref{99}. In this paper, it will not be necessary to compute $\phi_{(2)}$ and higher order corrections to $\phi$.\\

\noindent{\it Order $\epsilon^{0}$}: It is straightforward to show that the solution of \eqref{101} is 
\begin{equation}
\phi_{(0)}=ln\big(2(-\delta (u+u_{0}) +{\cal{D}})\big) \ .
\label{103}
\end{equation}
Note that this is precisely expression \eqref{100} written in the Gaussian normal coordinate $u$. This follows from the fact that for $L \rightarrow \infty$, $z/l$ and $u+u_{0}$ are identical.\\

\begin{figure}[ht]
    \centering
    \includegraphics[width=9cm]{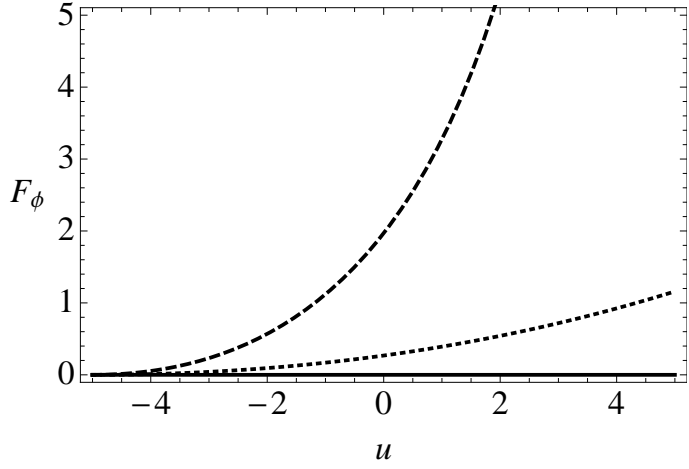}
    \caption{\label{fig:Fphi}
    Numerical solution for $F_{\phi}$ with ${\cal{C}}=2^{1/3}$, 
    ${\cal{D}}=1/2$ and $u_{0}=5$. We compute $F_{\phi}$ for $\delta = 0$ (solid), $\delta = 0.005$ (dotted) and $\delta = 0.01$ (dashed) respectively.}
\end{figure}

\noindent{\it Order $\epsilon^{1}$}: Since $\kappa_{(1)}\propto \hat{\kappa}_{(1)}(\sigma) $, we solve \eqref{102} using the ansatz
\begin{equation}
\phi_{(1)}=F_{\phi}(u)\hat{\kappa}_{(1)}(\sigma) \ .
\label{104}
\end{equation}
It follows that the $\sigma$ dependence cancels out and  \eqref{102} becomes a becomes a differential equation for $F_{\phi}(u)$ given by
\begin{equation}
F_{\phi}^{\prime\prime}-\frac{2\delta}{3(-\delta (u+u_{0})+{\cal{D}})}F_{\phi}^{\prime}-\frac{2\delta^{2}}{3(-\delta (u+u_{0})+{\cal{D}})^{2}}F_{\phi} 
-\frac{\delta}{{\cal{C}}(-\delta (u+u_{0})+{\cal{D}})^{\frac{4}{3}}}=0 \ .
\label{105}
\end{equation}
In determining this expression, we have used \eqref{98},\eqref{99} and \eqref{103}. The general solution of \eqref{105} is 
\begin{equation}
F_{\phi}(u)=-\frac{9}{4{ \cal{C}} \delta}({\cal{D}}-\delta (u+u_{0}))^{\frac{2}{3}}+C_{1}({\cal{D}}-\delta (u+u_{0}))^{-\frac{2}{3}}+C_{2}({\cal{D}}-\delta (u+u_{0})) 
\label{106} 
\end{equation}
where $C_{1},C_{2}$ are arbitrary constants. These constants can be completely specified by
noting that there is an additional constraint on $F_{\phi}$. This arises from the fact that, although being expressed in an $\epsilon$-expansion, the $\phi$ field is fixed to be the background heterotic solution \eqref{100}. Inserting \eqref{100}, \eqref{103} and \eqref{104} into
\begin{equation}
\phi(z)=\phi_{(0)}(u)+\epsilon\phi_{(1)}(u,\sigma)+{\cal{O}}(\epsilon^{2}) 
\label{107} 
\end{equation}
gives
\begin{equation}
\delta \frac{z}{l}=\delta (u+u_{0})+\epsilon(\delta (u+u_{0})-{\cal{D}})F_{\phi}\hat{\kappa}_{(1)}+{\cal{O}}(\epsilon^{2}) \ .
\label{108}
\end{equation}
Taking the limit $\delta \rightarrow 0$ while holding the kink width $l$ fixed, we get the additional constraint that
\begin{equation}
F_{\phi}\stackrel{\delta \rightarrow 0}{ \longrightarrow} 0 \ .
\label{109}
\end{equation}
Taylor expanding \eqref{106}, it is straightforward to show that this will be satisfied if and only if one chooses
\begin{equation}
C_{1}=\frac{9}{20}\frac{{\cal{D}}^{4/3}}{{\cal{C}}\delta} , ~\quad C_{2}=\frac{9}{5}\frac{{\cal{D}}^{-1/3}}{{\cal{C}}\delta}  \ .
\label{110}
\end{equation}
In this case, the first two terms in the Taylor expansion vanish and 
\begin{equation}
F_{\phi}=\delta f_{\phi} \ ,
\label{111}
\end{equation}
where
\begin{equation}
f_{\phi}=\frac{1}{2{\cal{C}}{\cal{D}}^{4/3}} (u+u_{0})^{2}+{\cal{O}}(\delta (u+u_{0})^{3})  ~ \ .
\label{112}
\end{equation}
Expression \eqref{106} with constants \eqref{110} can be numerically evaluated for given values of  ${\cal{C}}$, ${\cal{D}}$, $u_{0}$ and $\delta$. Several explicit examples are presented in Figure \ref{fig:Fphi}.\\

Finally, note that \eqref{108} can now be written as
\begin{equation}
\frac{z}{l}=(u+u_{0})-\epsilon ({\cal{D}}-\delta (u+u_{0})) f_{\phi}(u) {\hat{\kappa}}_{(1)}(\sigma)+{\cal{O}}(\epsilon^{2}) \ .
\label{113}
\end{equation}
It follows that the $\epsilon$-expansion of $\phi$ simply relates the original dimensionless coordinate $\frac{z}{l}$ to the Gaussian normal coordinate $u$ and worldvolume coordinates $\sigma^{\mu}$ order by order in 
$\epsilon$. For example, by definition $u=0$ on the kink hypersurface $\Sigma$. It follows that the location of the hypersurface in the original $\frac{z}{l}$ coordinate, evaluated to order $\epsilon$, is given by
\begin{equation}
\frac{z}{l}=u_{0}+\epsilon({\cal{D}}-\delta u_{0})f_{\phi}(u_{0}) {\hat{\kappa}}_{(1)}(\sigma) \ .
\label{113A}
\end{equation}
Note that, at this stage of the calculation, the $\sigma$-dependent factor $ {\hat{\kappa}}_{(1)}(\sigma)$ is necessarily undetermined. It can only be evaluated from the equation of motion of the worldvolume action--which we will determine below.
\\

\noindent {\it The $\chi$ Equation}-- ~Recall that the $\chi$ equation of motion in Gaussian normal coordinates is given by \eqref{66}. To order $\epsilon^{0}$ and $\epsilon^{1}$ in the $\epsilon$-expansion this becomes
\begin{eqnarray}
&&\chi_{(0)}^{\prime \prime}+\kappa_{(0)}\chi_{(0)}^{\prime}-\phi_{(0)}^{\prime}\chi_{(0)}^{\prime}-2(1-\chi_{(0)}^{2})(\frac{\delta}{3}-\chi_{(0)}e^{\phi_{(0)}})e^{-\phi_{(0)}}=0  \label{114} \\
&&\chi_{(1)}^{\prime \prime}+\kappa_{(0)}\chi_{(1)}^{\prime}+\kappa_{(1)}\chi_{(0)}^{\prime}-\phi_{(1)}^{\prime}\chi_{(0)}^{\prime}-\phi_{(0)}^{\prime}\chi_{(1)}^{\prime}+\frac{2}{3}e^{-2\phi_{(0)}}\big(2\delta\chi_{(0)}+3e^{\phi_{(0)}}  \nonumber \\
&&(1-3\chi_{(0)}^{2} )\big) \chi_{(1)}
+\frac{2}{3}e^{-2\phi_{(0)}}\phi_{(1)}(1-\chi_{(0)}^{2})(2\delta-3e^{2\phi_{(0)}}\chi_{(0)})=0 \label{115} 
\end{eqnarray}
respectively, where $\kappa_{(0)}$,$\kappa_{(1)}$ are given in \eqref{98},\eqref{99} and $\phi_{(0)}$,$\phi_{(1)}$ in \eqref{103},\eqref{104}. It will not be necessary to compute $\chi_{(2)}$ and higher order corrections to $\chi$.\\

\noindent{\it Order $\epsilon^{0}$}: It is straightforward to show that the solution of \eqref{114} is 
\begin{equation}
\chi_{(0)}=tanh(u) \ .
\label{116}
\end{equation}
Note that this is precisely expression \eqref{37} written in the Gaussian normal coordinate $u$ with $v=1$.  This follows from the fact that for $L \rightarrow \infty$, $z/l$ and $u+u_{0}$ are identical.\\

\noindent{\it Order $\epsilon^{1}$}: Since $\kappa_{(1)}, \phi_{(1)}\propto \hat{\kappa}_{(1)}(\sigma) $, we solve \eqref{115} using the ansatz
\begin{equation}
\chi_{(1)}=F_{\chi}(u)\hat{\kappa}_{(1)}(\sigma) \ .
\label{117}
\end{equation}
It follows that the $\sigma$ dependence cancels out and  \eqref{115} becomes a becomes a differential equation for $F_{\chi}(u)$ given by
\begin{eqnarray}
&&F_\chi'' -\big(\frac{2\delta{\cal{C}}^{3}}{3}(e^{-2A})^{3}+\phi_{(0)}^{\prime}\big)F_\chi'+\frac{2}{3}e^{-2\phi_{(0)}}\big(2\delta\chi_{(0)}+3e^{\phi_{(0)}} (1-3\chi_{(0)}^{2} )\big) F_\chi \nonumber \\
&&
+\frac{2}{3}e^{-2\phi_{(0)}}(1-\chi_{(0)}^{2})(2\delta-3e^{2\phi_{(0)}}\chi_{(0)})F_\phi +(e^{-2A}-F_\phi')\chi_{(0)}^{\prime}=0  \nonumber\\
\label{118} 
\end{eqnarray}
Unlike the case for $F_{\phi}$, we can only solve the $F_{\chi}$ equation numerically for given values of ${\cal{C}}$, ${\cal{D}}$, $u_{0}$ and $\delta$. Several explicit examples are presented in Figure 2.\\

\begin{figure}[ht]
    \centering
    \begin{tabular}{ccc}
    \includegraphics[width=4.5cm]{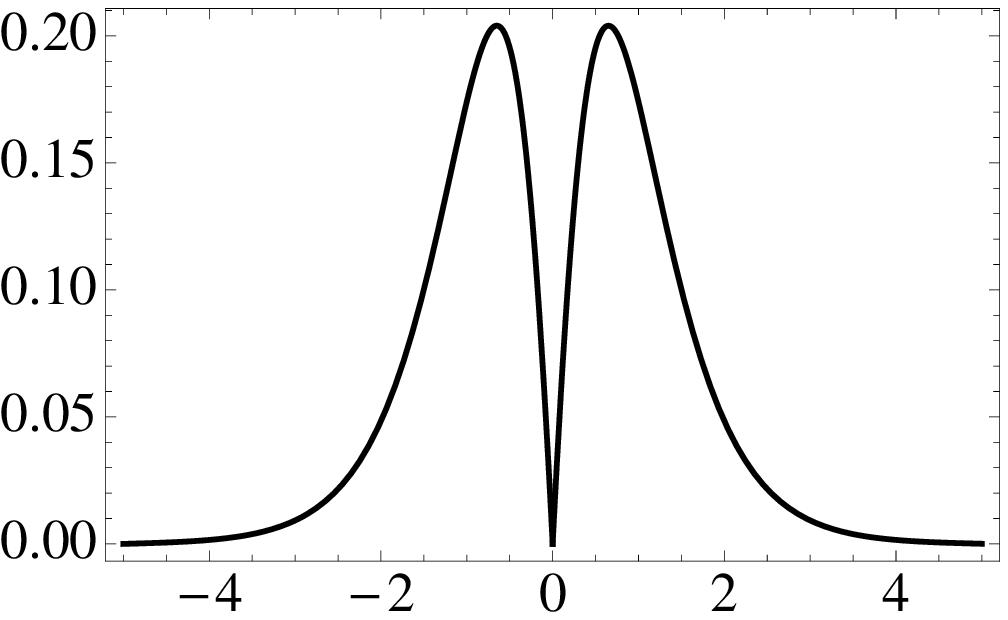}
    &
    \includegraphics[width=4.5cm]{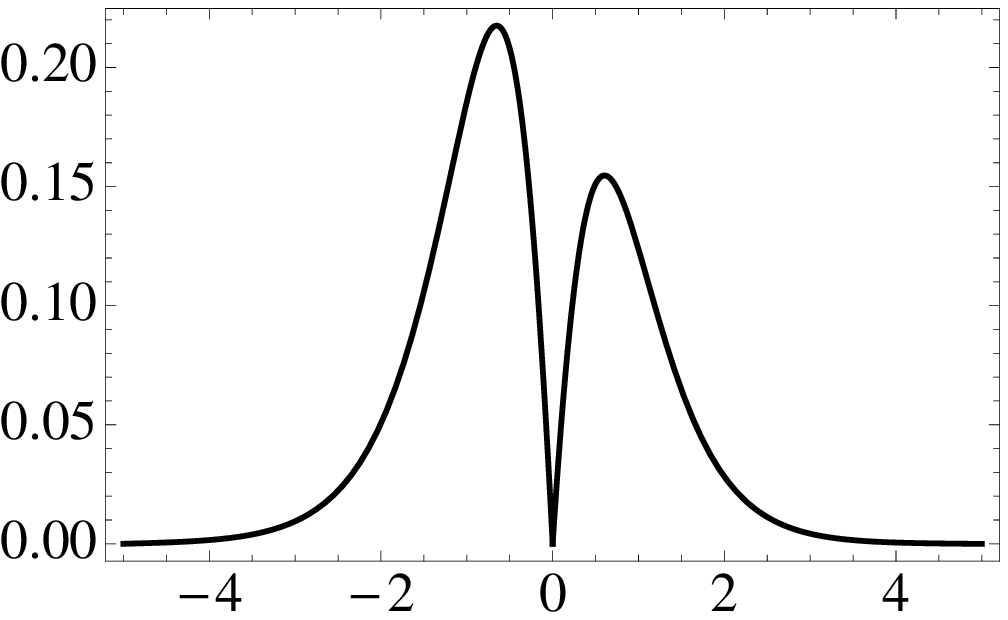}
    &
    \includegraphics[width=4.5cm]{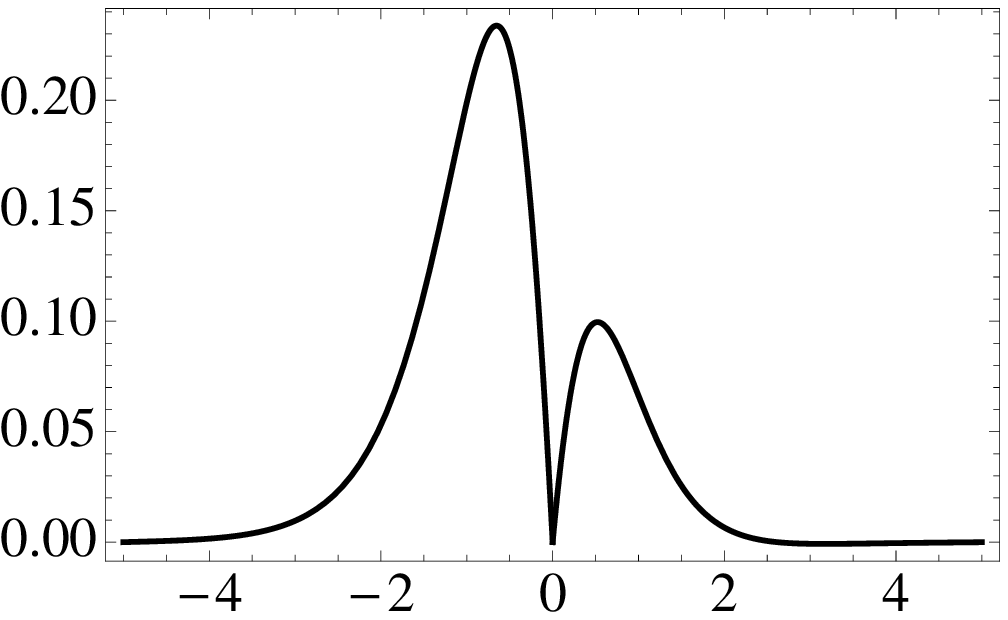}
    \end{tabular}
    \caption{\label{fig:Fchi} Numerical solution for $F_\chi$ with ${\cal{C}}=2^{1/3}$, ${\cal{D}}=1/2$, $u_{0}=5$ and, from left to right, $\delta = 0,~0.005,~0.01$. The x- and y-axis correspond to $u$ and $F_{\chi}$ respectively.}
\end{figure}
\noindent We conclude that to this order in $\epsilon$,
\begin{equation}
\chi=tanh(u)+\epsilon F_{\chi}(u)\hat{\kappa}_{(1)}(\sigma)+{\cal{O}}(\epsilon^{2}) 
\label{119}
\end{equation}
is kink soliton that depends on the four spacetime coordinates $\sigma^{\mu}$ as well as on $u$.
As in the flat spacetime and AdS cases discussed in \cite{KOS}, $\chi$ is continuous, but not continuously differentiable, across the $u=0$ hypersurface.

\section{The Worldvolume Effective Action}

The worldvolume effective action for the kink soliton in $\chi$ can now be calculated to any required accuracy in the $\epsilon$ expansion. It is given by
\begin{equation} 
S_{4}=\int_{M_{4}}{{\rm d}^{4}\sigma \sqrt{-h}|_{u=0}} {\hat{\cal{L}}}_{\chi}
\label{120}
\end{equation}
where
\begin{equation}
{\hat{\cal{L}}}_{\chi}=\int{{\rm d}z_{g}} J {\cal{L}}_{\chi}\,, \quad   J= \frac{\sqrt{-g}}{\sqrt{-h}|_{u=0}}\ .
\label{121}
\end{equation}
${\cal{L}}_{\chi}$ is the decoupled $\chi$ Lagrangian density \eqref{49}.
It convenient to write $z_{g}=lu$ and let
\begin{equation}
{\cal{L}}_{\chi}=-\frac{\eta^2}{4l^2}{\cal{L}} \ ,
\label{121A}
\end{equation}
where ${\cal{L}}$ is dimensionless. Then
\begin{equation}
{\hat{\cal{L}}}_{\chi}=-\frac{\eta^2}{4l}\int{{\rm d}u} J {\cal{L}}\ .
\label{121AA}
\end{equation}

First consider the factor $J$. The metric $g_{mn}$ is the background heterotic metric \eqref{13A},\eqref{22C} in $B=0$ gauge. Taylor expanding $\sqrt{-g}$ around the location of the kink hypersurface, that is, $u=0$ in Gaussian normal coordinates, 
we find 
\begin{equation}
\sqrt{-g} = \sqrt{-h}|_{u=0} \big( 1 + \epsilon u\kappa|_{u=0} + \frac{\epsilon^{2}}{2} u^2(\kappa^2 - \kappa_{mn}\kappa^{mn} - L^{2}R^{(5)}_{mn}n^m n^n)|_{u=0} +\dots\big)
\label{121C}
\end{equation}
Using \eqref{53},\eqref{58} and $h_{mn}h^{mn}=4$, gives
\begin{equation}
L^{2}R^{(5)}_{mn}n^m n^n=L^{2}R^{(5)}_{rspq}n^{s}n^{q}h^{r}_{m}h^{p}_{n}h^{mn}
=\frac{5}{9\left(\mathcal{D} - \delta (u+u_0)\right)^2}(\delta/\epsilon)^2 \ .
\label{121D}
\end{equation}
It follows that
\begin{equation}
J=1+\epsilon J_{(1)} +\frac{\epsilon^{2}}{2}J_{(2)}+{\cal{O}}(\epsilon^{3}) \ ,
\label{122}
\end{equation}
where
\begin{equation}
J_{(1)}= u \kappa|_{u=0}\,, \quad J_{(2)}= u^{2}\Big(\kappa^{2}-\kappa_{mn} \kappa^{mn} - \frac{5}{9\left(\mathcal{D} - \delta (u+u_0)\right)^2}(\delta/\epsilon)^2\Big)|_{u=0} \ .
\label{123}
\end{equation}

Next consider the dimensionless Lagrangian density. Neglecting the spatial derivative terms, which are higher-order in $\epsilon$ than required in this paper, ${\cal{L}}$ is given by
\begin{equation}
{\cal{L}}= e^{-\phi}(\chi')^{2} + \frac{4}{3} \delta e^{-2\phi} (\chi - \chi^3/3) + e^{-\phi} (1-\chi^2)^2.
\label{121B}
\end{equation}
Inserting the $\epsilon$ expansions of the background dilaton \eqref{107} and the $\chi$ kink solution \eqref{119} into this expression,
it follows that
\begin{equation}
{\cal{L}}={\cal{L}}_{(0)}+\epsilon {\cal{L}}_{(1)}+\frac{\epsilon^{2}}{2}{\cal{L}}_{(2)}+{\cal{O}} (\epsilon^{3}) \ .
\label{124}
\end{equation}
The component Lagrangians $ {\cal{L}}_{(i)}$ are complicated functions of $\phi_{(0)}$,$\phi_{(1)}$,$\chi_{(0)}$ and $\chi_{(1)}$. For example,
\begin{equation}
{\cal{L}}_{(0)} = e^{-\phi_{(0)}}(\chi_{(0)}')^2 + \frac{4}{3} \delta e^{-2\phi_{(0)}} (\chi_{(0)} - \chi_{(0)}^3/3) + e^{-\phi_{(0)}} (1-\chi_{(0)}^2)^2 \ ,
\label{125}
\end{equation}
that is, the original Lagrangian evaluated on the field configurations. 

Before presenting the other component Lagrangians, however, we simplify the calculation in two ways. First, recall that the expressions for $\phi_{(0)}$,$\phi_{(1)}$ and $\chi_{(1)}$, given in \eqref{103},\eqref{104}, and \eqref{117} respectively, contain the arbitrary constants ${\cal{C}}$ and ${\cal{D}}$. In the case of  $\phi_{(1)}$ and $\chi_{(1)}$, they enter through the functions $F_{\phi}$ in \eqref{106},\eqref{110} and $F_{\chi}$ in \eqref{118}. We will henceforth, for specificity, make the canonical choices 
\begin{equation}
{\cal{C}}=2^{1/3}~, \quad {\cal{D}}=\frac{1}{2} 
\label{126}
\end{equation}
as first discussed in \eqref{22F} and used in calculating Figures 1 and 2. Second, in addition to the $\epsilon$ expansion in \eqref{124} we will also do a systematic power series expansion in the small parameter $\delta$.  To avoid having to compute $\phi_{(2)}$ and $\chi_{(2)}$, we work to second order in the expansion parameters only. The $\delta$ parameter enters in two ways--through the Lagrangian density \eqref{121B} itself and in the expressions for $\phi_{(0)}$,$\phi_{(1)}$ and $\chi_{(1)}$. The function $\phi_{(0)}$ appears in ${\cal{L}}_{(0)}$ and, hence, one must Taylor expand \eqref{103} to order $\delta^{2}$. The $\delta$ dependence of $\phi_{(1)}$ is contained in the function $F_{\phi}$, and recall that $F_{\phi}\propto \delta$. Since $\phi_{(1)}$ does not occur in ${\cal{L}}_{(0)}$, we need only its lowest order value. 
Using \eqref{111},\eqref{112} and \eqref{126}, we find to lowest order that
\begin{equation}
F_{\phi}= \delta(u+u_0)^2.
\label{126A}
\end{equation}
The $\delta$ dependence of $\chi_{(1)}$ is contained in the function $F_{\chi}$, which begins at order $\delta^{0}$. Hence, we will need the first two orders in its $\delta$-expansion.
For this purpose, it will be useful to write
\begin{equation}
	F_\chi = F_\chi^{(0)} + \delta F_{\chi}^{(1)} + \mathcal{O}(\delta^2) \ ,
\label{126B}
\end{equation}
where, for example, $F_\chi^{(0)}$ is the solution to the $F_\chi$ differential equation \eqref{118} in the $\delta \to 0$ limit with $\mathcal{C}$ and $\mathcal{D}$ set to their canonical values. Using the equations of motion \eqref{101},\eqref{102} and \eqref{114},\eqref{115} for $\phi_{(0)}$,$\phi_{(1)}$ and $\chi_{(0)}$, $\chi_{(1)}$ respectively, and then expanding these fields as a power series in $\delta$, each ${\cal{L}}_{(i)}$ splits into its flat-space value plus terms which depend on $\delta$. Explicitly, to ${\cal{O}}(\epsilon^3,\epsilon^2\delta,\epsilon\delta^2,\delta^3)$  we find that
\begin{equation}
{\cal{L}}_{(i)}= {\cal{L}}_{(i){\rm flat}} + \Delta_{(i)} \ ,
\label{127}
\end{equation}
where
\begin{eqnarray}
&&\quad {\cal{L}}_{(0)\rm flat}=2\chi_{(0)}^{\prime 2}, \quad {\cal{L}}_{(1)\rm flat}=2 (\chi_{(0)}^{\prime}\chi_{(1)\rm flat})^{\prime} \label{128} \\
&& {\cal{L}}_{(2) \rm flat}= 2 \left( (\chi_{(0)}^{\prime}\chi_{(2))})^{\prime}+ (\chi_{(1)\rm flat}^{\prime}\chi_{(1)\rm flat})^{\prime} +{\hat{\kappa}}_{(1)}\chi_{(0)}^{\prime}\chi_{(1)\rm flat} \right) \nonumber
\end{eqnarray}
and we have defined $\chi_{(1)\rm flat} = F_\chi^{(0)}\hat{\kappa}_{(1)}$. Note that it is not necessary to expand $\chi_{(2)}$ since this only appears as a total derivative term which we will drop. To the order we are working, the $\delta$-dependent terms are given by
\begin{eqnarray}
&&\Delta_{(0)} = \delta\Delta_{(0)}^{(1)} + \delta^2\Delta_{(0)}^{(2)} + \mathcal{O}(\delta^3) \label{129}\\
&&\Delta_{(1)} = \left[\delta \Delta_{(1)}^{(1)} + \mathcal{O}(\delta^2) \right] \hat{\kappa}_{(1)} \label{130}\\
&&\Delta_{(2)} = \mathcal{O}(\delta) \label{131}
\end{eqnarray}
with
\begin{eqnarray}		
	&&\Delta_{(0)}^{(1)}
		 = -\frac{4}{9}\chi_{(0)}(\chi_{(0)}^2 - 3) + 4(u+u_0)\chi_{(0)}'^2 \label{132} \\
         && \Delta_{(0)}^{(2)}
		 = \frac{8}{9}(u+u_0)\left(6\chi_{(0)} - 2\chi_{(0)}^3 + 9(u+u_0)\chi_{(0)}'^2\right) \label{133}\\
	&&\Delta_{(1)}^{(1)}
		 = 2(\chi_{(0)}'F_\chi^{(1)})'-(u+u_0)^2\big(\chi_{(0)}'^2 + (\chi_{(0)}^2 - 1)^2\big) \\
		 &&\qquad\quad+ \frac{4}{3}\big(F_\chi^{(0)}+ 3(u+u_0)(F_\chi^{(0)})'\big)\chi_{(0)}'. \label{134}
\end{eqnarray}
Note that we have not presented the formula for $\Delta_{(2)}$ since it contributes an $\mathcal{O}(\epsilon^2\delta)$ term to the effective Lagrangian. 

At this point, we introduce two additional simplifications. First, note that the parameter $u_{0}=z_{0}/l$ enters all but the ${\cal{L}}_{(0) \rm flat}$ term in \eqref{127}. Recalling that the $z$ parameter has the range $0 \leq z \leq \pi \rho$, we now, for simplicity, choose the kink hypersurface to be at the center of this interval in the $\epsilon\rightarrow 0$ limit. That is, we take
\begin{equation}
u_{0}=\frac{\pi\rho}{2l} \ .
\label{135}
\end{equation}
It follows that the Gaussian normal coordinate $u$ lies in the symmetric range $-\pi\rho/2l \leq u \leq \pi\rho/2l$. It is important to note from \eqref{22B},\eqref{22C} and the fact that 
$z \in [0,\pi\rho]$ that choosing ${\cal{D}}=1/2$ restricts $1/\alpha>2\pi\rho$ and, hence,
\begin{equation}
\delta<\frac{1}{4u_{0}} \ .
\label{135A}
\end{equation}
We will impose this restriction henceforth. Second, since we are expanding in the small parameter $\epsilon$, the fluctuation length $L$ of the kink hypersurface should be large compared to the kink width. It is natural, therefore, to simply take $L>\pi\rho/2$ and, hence, 
\begin{equation}
\epsilon<\frac{1}{u_{0}} \ .
\label{135B}
\end{equation}
It follows from these two conditions that, since $1/\alpha$ and $L$ lie outside the heterotic interval,  the boundary walls at $\pm u_{0}$ will act as a natural cut-off for any integrals appearing in the calculation.

Multiplying $J$ and $\mathcal{L}$ gives
\begin{eqnarray}
&&J{\cal{L}} = {\cal{L}}_{\rm flat (0)}\left(1+\frac{\epsilon^{2}}{2}J_{(2)}  \right)+\epsilon {\cal{L}}_{\rm flat (1)}\left(1+\epsilon J_{(1)}  \right)+\frac{\epsilon^{2}}{2}{\cal{L}}_{\rm flat (2)} \label{136} \\
&&\quad + \Delta_{(0)}\left(1+ \epsilon J_{(1)} + \frac{\epsilon^2}{2}J_{(2)}\right) + \epsilon \Delta_{(1)}(1 + \epsilon J_{(1)}) + \frac{\epsilon^2}{2} \Delta_{(2)} + \mathcal{O}(\epsilon^3) \nonumber
\end{eqnarray}
Note that we have dropped $\epsilon {\cal{L}}_{\rm flat (0)} J_{(1)}$ since it is odd in $u$ and thus integrates to zero over the symmetric integration region. In order to relate the wall extrinsic curvature $\kappa|_{u=0}$ to the function $\hat{\kappa}_{(1)}$ appearing in $\Delta_{(1)}$, we substitute the expansions for $h^{mn}$ and $\kappa_{mn}$ into \eqref{96}. Expanding in $\delta$ and evaluating at $u = 0$, we find the relation
\begin{equation}
	\hat{\kappa}_{(1)}  = \kappa|_{u=0} + \frac{4\delta}{3\epsilon} + \mathcal{O}(\delta) \ .
\label{137}
\end{equation}
Inserting the expressions for $\Delta_{(i)}$ and writing $\hat{\kappa}_{(1)}$ in terms of $\kappa|_{u=0}$, we find that
\begin{align}
	J\mathcal{L}
		& = 2\chi_{(0)}'^2 + \epsilon^2 \chi_{(0)}'^2 u^2 \Big(\kappa^{2}-\kappa_{mn} \kappa^{mn} - \frac{20}{9}(\delta / \epsilon)^2\Big)|_{u=0}  + \nonumber \\
		& \quad - \epsilon^2 \kappa|_{u=0}^2 \chi_{(0)}' F^{(0)}_\chi + \frac{16}{9}\delta^2\chi_{(0)}' F^{(0)}_\chi + \delta \Delta_{(0)}^{(1)} + \epsilon \kappa|_{u=0} u \delta \Delta_{(0)}^{(1)} + \label{138}\\
		& \quad + \delta^2 \Delta_{(0)}^{(2)} + \epsilon \kappa|_{u=0} \delta \Delta_{(1)}^{(1)} + \frac{4}{3}\delta^2 \Delta_{(1)}^{(1)} + \mathcal{O}(\epsilon^3,\epsilon^2\delta,\epsilon\delta^2,\delta^3) \nonumber
\end{align}
up to total derivative terms such as $\epsilon \mathcal{L}_{(1)\rm flat}$ and $2(\chi_{(0)}'F_\chi^{(1)})'$ among others. These total derivatives integrate to boundary terms which are rapidly decaying functions of $u$. Thus, provided we take the boundary walls to be at a sufficiently large distance from the kink hypersurface in units of $l$, such terms are extremely small and can be ignored.
Defining 
\begin{equation}
\hat{\kappa}(\sigma) = \kappa |_{u=0} \ , 
\label{139}
\end{equation}
the effective Lagrangian is then given by
\begin{equation}
	\hat{\mathcal{L}}_\chi 
		= \hat{\mathcal{L}}_{(0)} + \epsilon \hat{\mathcal{L}}_{(1)} + \frac{\epsilon^2}{2}\hat{\mathcal{L}}_{(2)}+{\cal{O}}(\epsilon^{3},\epsilon^{2}\delta,\epsilon\delta^{2},\delta^{3}) \ ,
\label{140}
\end{equation}
where
\begin{align}
	\hat{\mathcal{L}}_{(0)}
		& = -\frac{\eta^2}{2l}\left[I_I+2\delta u_0 I_I +\delta^{2}\left(\frac{8}{9}  I_{III} +  \frac{1}{2}I_{IV} + \frac{2}{3} I_{V}\right)\right] \label{141} \\
	\hat{\mathcal{L}}_{(1)}
		& = -\frac{\eta^2}{4l}\hat{\kappa} \delta (I_{VI} + I_{V}) \label{142}\\
	\hat{\mathcal{L}}_{(2)}
		& = -\frac{\eta^2}{2l}  \left[ I_{II}(\hat{\kappa}^{2}-\hat{\kappa}_{mn} \hat{\kappa}^{mn} - (20/9)(\delta/\epsilon)^2) - \hat{\kappa}^2 I_{III} \right] \ . \label{143}
\end{align}
The coefficients are 
\begin{align}		
	&I_I
		 = \int du \, \chi_{(0)}'^2 = \frac{4}{3} \label{144}\\
	&I_{II}
		 = \int du \, u^2 \chi_{(0)}'^2 = \frac{\pi^2-6}{9}\label{145} \\
	&I_{III}
		 = \int du \, F_\chi \chi_{(0)}' = \frac{5}{18} \label{146} \\
	&I_{IV}
		 = \int du \, 	\Delta_{(0)}^{(2)}|_{\rm even} = \frac{8}{9}(\pi^2-4+12u_0^2) + \frac{32}{9}\int du \, u \tanh u \label{147}\\
	&I_{V}
		 = \int du \, \Delta_{(1)}^{(1)} = \frac{1}{27}(13 - 3\pi^2 - 72 u_0^2) \label{148} \\
	&I_{VI}
		 = \int du \, u \Delta_{(0)}^{(1)}|_{\rm odd} = \frac{4}{9}(\pi^2 - 4) + \frac{1}{9}\int du \, u\sech^4 u \sinh 4 u \label{149}
\end{align}
with
\begin{eqnarray}
	&&\Delta_{(0)}^{(1)}|_{\rm odd}
		 = \frac{1}{9}\sech^4 u (36 u + 4 \sinh 2u + \sinh 4u) \label{150}\\
	&&\Delta_{(0)}^{(2)}|_{\rm even}
		 = 8(u^2+u_0^2)\sech^4 u + \frac{16}{9} u (2 + \sech^2 u)\tanh u \label{151}\\
	&&\Delta_{(1)}^{(1)}
		 = -2(u+u_0)^2 \sech^4 u + \frac{4}{3}\sech^2 u \, \big(F^{(0)}_\chi(u) \label{152}\\
	&&\qquad\qquad\qquad\qquad\qquad\qquad~~~ ~ +3(u+u_0) F^{(0)\prime}_{\chi}(u)\big)\nonumber 
\end{eqnarray}
and $u_{0}$ the arbitrary dimensionless parameter given in \eqref{135}. As discussed above, it follows from the assumptions \eqref{135A}, \eqref{135B} that all integrals are over $(-u_{0},+u_{0})$. For $u_{0}$ sufficiently large, the coefficients $I_{I}$, $I_{II}$, $I_{III}$ and $I_{V}$ are well-approximated by integrating over $(-\infty,+\infty)$. The results are presented above. However, this is not the case for the $I_{IV}$, $I_{VI}$ coefficients since each contains, in addition to a finite piece,  a primitively divergent integral. Cutting these off at $\pm u_{0}$, we find
\begin{equation}
I_{IV}=\frac{8}{9}(\pi^2-4+12u_0^2) + \frac{32}{9}u_0^2 , \quad I_{VI}=\frac{4}{9}(\pi^2 - 4) +\frac{8}{9} u_0^2 \ .
\label{152A}
\end{equation}
The final terms in $I_{IV}$ and $I_{VI}$ arise from the cut-off integral in \eqref{147} and \eqref{149} respectively.

Before stating the final result, we note that the Lagrangian can be simplified by writing it in terms of the intrinsic scalar curvature using the Gauss-Codazzi relation
\begin{equation}
\hat{R}^{(4)}={\hat{K}}^{2}-{\hat{K}}^{n}_{m} {\hat{K}}^{m}_{n} + R^{(5)} - 2R^{(5)}_{mn}n^m n^n.
\label{156}
\end{equation}
It follows from the metric \eqref{13A},\eqref{22C} that the five-dimensional  Ricci tensor for the heterotic background geometry in $B=0$ gauge is
\begin{equation}
R^{(5)} = -4(5A'^2 + 2A'') = \frac{7\alpha^2}{9(\mathcal{D}-z\alpha)^2} \ .
\label{157}
\end{equation}
Using \eqref{113} to relate the $B=0$ gauge coordinate $z$ to the dimensionless Gaussian normal coordinate $u$, Taylor expanding in both $\epsilon$ and $\delta$ to the required order, as well as setting ${\cal{D}}=1/2$ gives
\begin{equation}
L^2 R^{(5)} = \frac{28}{9} \left( \frac{\delta}{\epsilon} \right)^2 + \mathcal{O}(\epsilon,\delta^{3}) \ .
\label{158}
\end{equation}
Similarly, from \eqref{121D} we find 
\begin{equation}
L^{2}R^{(5)}_{mn}n^m n^n=\frac{20}{9} \left( \frac{\delta}{\epsilon} \right)^2 + \mathcal{O}(\delta^{3}) \ .
\label{158A}
\end{equation}
Putting this together, it follows that to the required order
\begin{equation}
\hat{R}^{(4)}={\hat{K}}^{2}-{\hat{K}}^{n}_{m} {\hat{K}}^{m}_{n}-\frac{4}{3L^{2}} \left( \frac{\delta}{\epsilon} \right)^2 + \mathcal{O}(\epsilon,\delta^{3}) \ .
\label{158B}
\end{equation}

Rewritten in dimensionful variables, using \eqref{158B} and truncating the action at second order in the parameters, the worldvolume Lagrangian \eqref{140} is given by
\begin{equation}
\hat{{\cal{L}}}_{\chi}=-\frac{2\eta^2}{3l}{\cal{M}}_{0}^{4}\Big(1+C_{0}\hat{K} +C_{I}{\hat{R}}^{(4)}+C_{II}{\hat{K}}^{2} \Big) \ ,
\label{155}
\end{equation}
where
\begin{equation}
{\cal{M}}_{0}^{4}= 1+2\delta u_0 +\delta^{2}\left(-\frac{1}{3}I_{II} + \frac{2}{3}  I_{III} +  \frac{3}{8}I_{IV} + \frac{1}{2} I_{V}\right)
\label{159}
\end{equation}
and 
\begin{equation}
C_{0}= \Big(\frac{I_V+I_{VI}}{{\cal{M}}_{0}^{4}}\Big)\frac{3l}{8}\delta,~~~C_{I}= \Big(\frac{I_{II}}{{\cal{M}}_{0}^{4}}\Big) \frac{3l^2}{8},~~~C_{II}= -\Big(\frac{I_{III}}{{\cal{M}}_{0}^{4}}\Big)\frac{3l^2}{8} \ .
\label{157} 
\end{equation}
Inserting the values of the $I_{II}$, $I_{III}$, $I_{V}$ coefficients given in \eqref{145}, \eqref{146},\eqref{148} and the $I_{IV}$, $I_{VI}$ coefficients in \eqref{152A}, it follows that
\begin{equation}
{\cal{M}}_{0}^{4}= 1+2\delta u_{0}+\delta^{2} \left[\frac{13}{54}\left(\pi^{2}-\frac{37}{13}\right)+4u_{0}^{2}  \right]
\label{157A}
\end{equation}
and
\begin{equation}
C_{0}= \frac{1}{8}\Big(\pi^{2}-\frac{35}{9}-\frac{16}{3}u_{0}^{2}\Big)\frac{l \delta}{{\cal{M}}_{0}^{4}},~~C_{I}= \Big(\frac{\pi^{2}-6}{24}\Big) \frac{l^2}{{\cal{M}}_{0}^{4}},~~C_{II}= -\frac{5}{48}\frac{l^2}{{{\cal{M}}_{0}^{4}}} \ .
\label{157B} 
\end{equation}
Note that in the limit $\alpha \rightarrow 0$ and, hence, $\delta \rightarrow 0$, Lagrangian \eqref{155} becomes the flat spacetime Lagrangian presented in \cite{KOS}. For a specified value of $u_{0}$, the overall coefficient ${\cal{M}}_{0}^{4}$ as well as the three coupling parameters $C_{0}/l$, $C_{I}/l^{2}$ and $C_{II}/l^{2}$ can be calculated numerically as functions of $\delta$. For example, these parameters are plotted in graphs (A),(B),(C) and (D) respectively in Figure 3 for $u_{0}=5$. Note that $C_{0}$ has the correct limiting value of $C_{0} \rightarrow 0$ as $\delta \rightarrow 0$.

\begin{figure}[H]
    \centering
    \begin{tabular}{cc}
    \includegraphics[width=6cm]{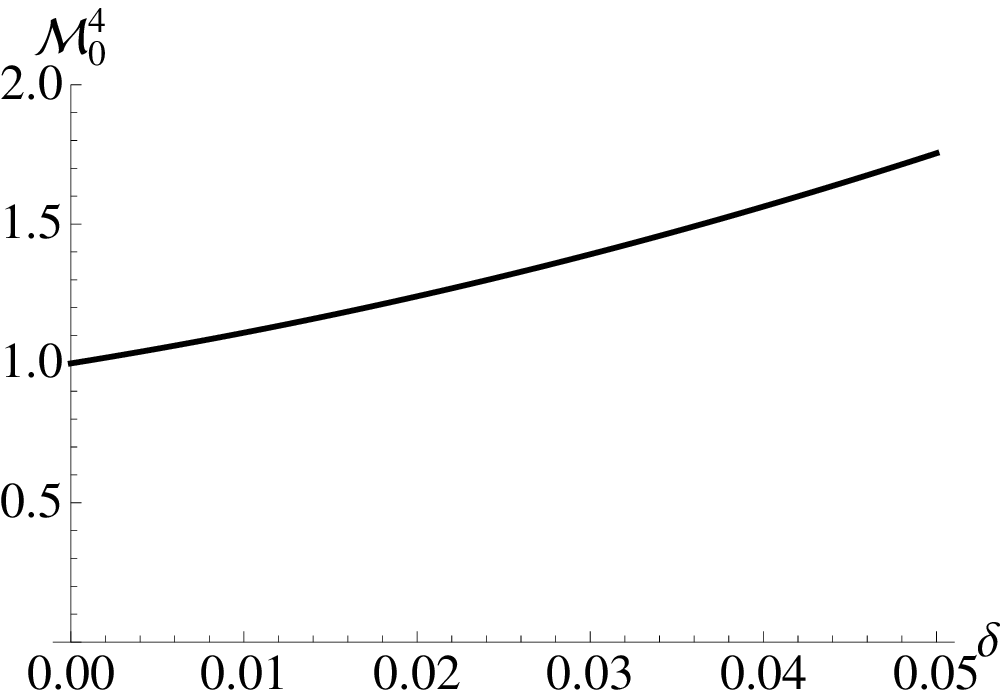}{(A)} & \includegraphics[width=6cm]{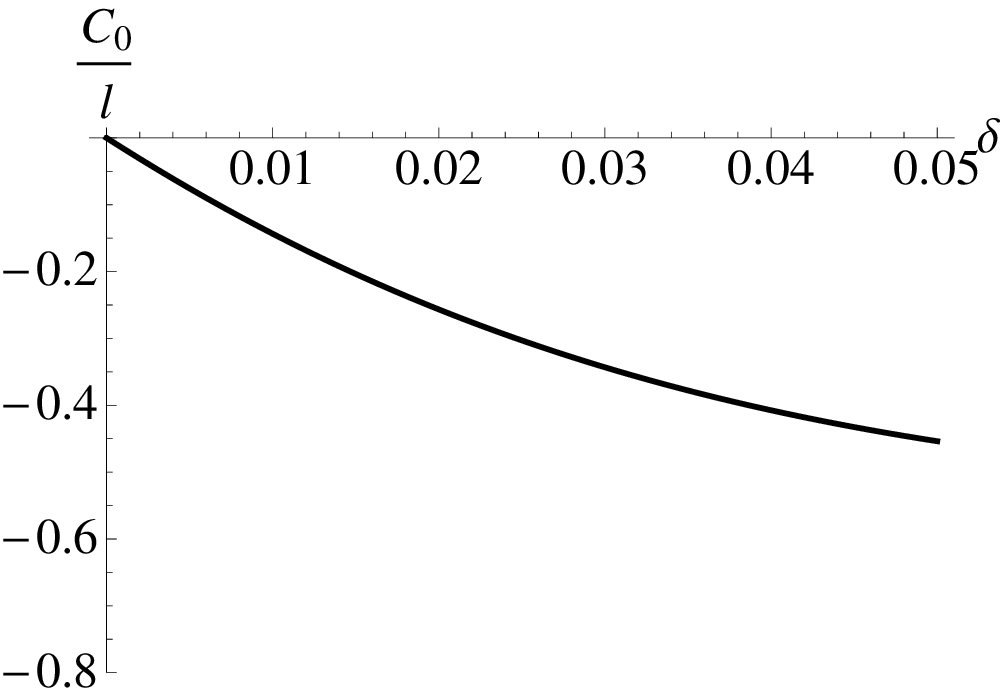}(B) \\
    \includegraphics[width=6cm]{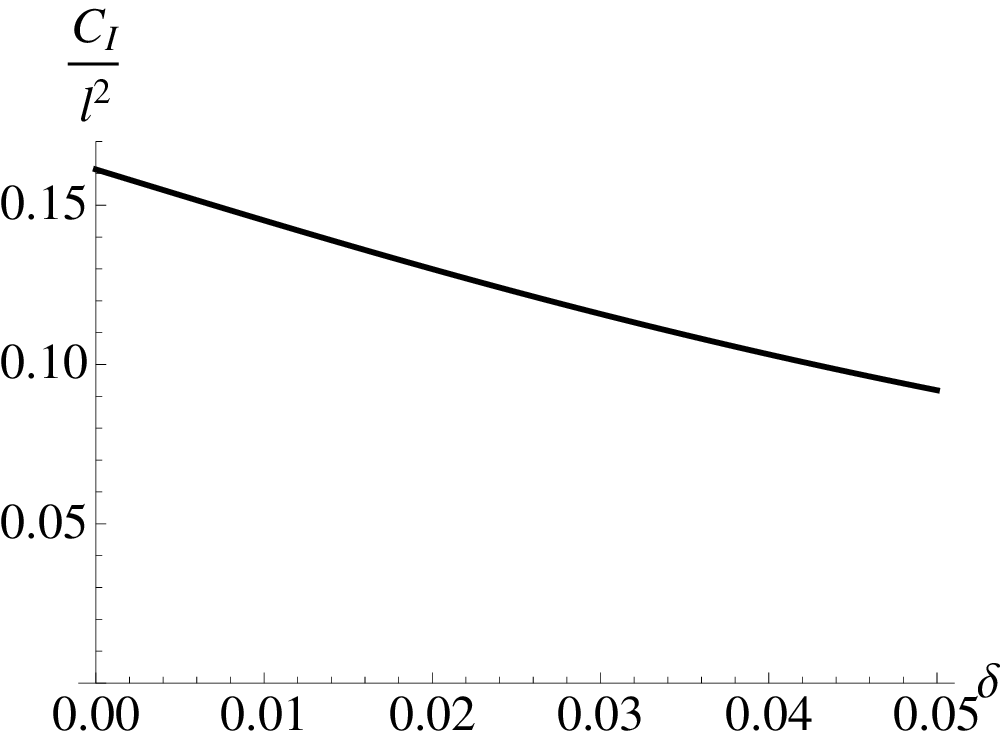}(C) & \includegraphics[width=6cm]{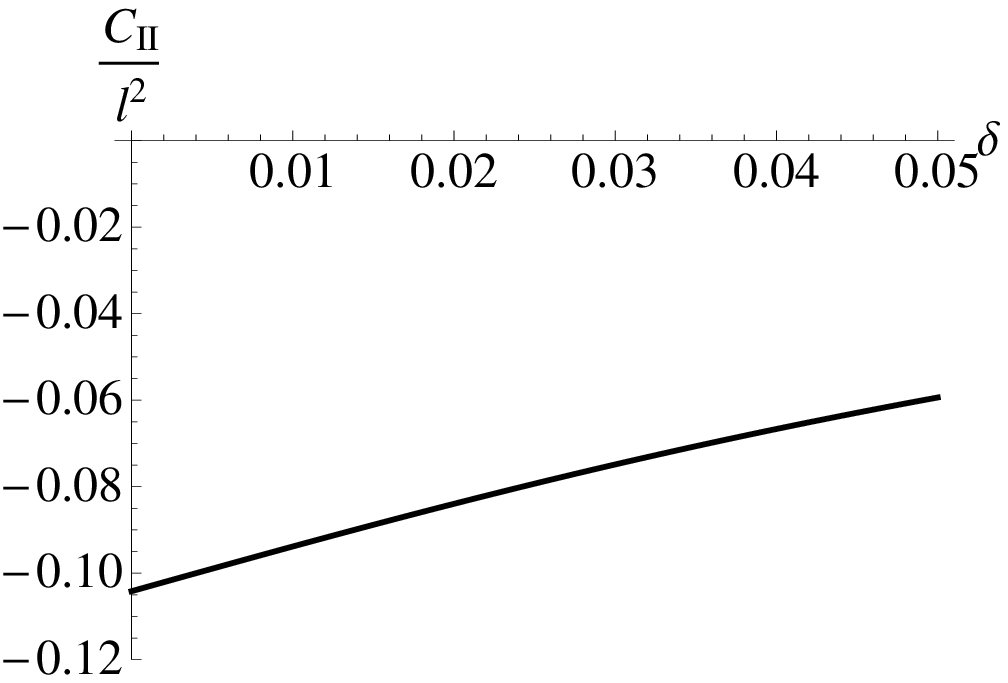}(D) \\
    \end{tabular}      
   \caption{\label{fig:C} Plots of $\mathcal{M}_0$, 
    $C_0 / l$, $C_{I} / l^{2}$ and $C_{II} / l^{2}$ as functions of $\delta$ for $u_0 = 5$. Note that $ 0\leq \delta < 1/4u_0=0.05$.}
\end{figure}

\begin{figure}[ht]
    \centering
    \includegraphics[width=9cm]{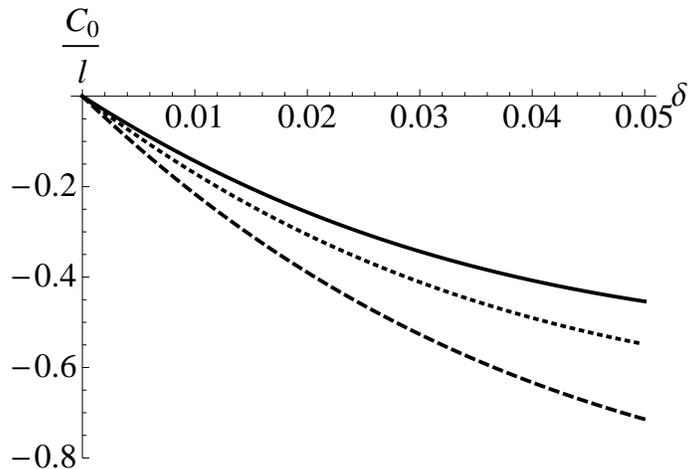}
    \caption{\label{fig:C0}
    Plots of the $C_0/l$ for the natural choice of $\epsilon > 1/u_0 = 0.2$  (solid), for $\epsilon = 0.25$ (dotted) and for the upper bound on a perturbative expansion $\epsilon = 1$ (dashed).}
\end{figure}

For completeness, we consider the implications of relaxing assumption \eqref{135B}. In this case, the natural cut-off in units of $l$ is $1/\epsilon$. Repeating the above calculation with the new cut-off, we find that $\mathcal{M}_0$, $C_I$ and $C_{II}$ are almost unaffected whereas $C_0$ can receive sizable corrections. These are displayed for various values of $\epsilon$ in Figure \ref{fig:C0}.

Finally, we note that all $\sigma^{\mu}$-dependent quantities in the effective action--that is, $\sqrt{-h}|_{u=0}$, ${\hat{K}}$, ${\hat{R}}^{(4)}$ and ${\hat{K}}^{2}$--can be expressed in terms of a worldvolume scalar field $\pi(\sigma)$ corresponding to translation in the fifth direction in heterotic spacetime. The explicit formulas are presented in the Appendix.


\section*{Acknowledgments}
We would like to thank Andre Lukas for bringing his work on heterotic kinks to our attention, as well as Kurt Hinterbichler and Justin Khoury for many enlightening discussions. The work of Burt Ovrut  is supported in part by the DOE under contract No. DE-AC02-76-ER-03071.  B.A.O. acknowledges partial support from the NSF RTG grant DMS-0636606 and from NSF Grant 555913/14 for International Collaboration. James Stokes acknowledges support by NASA ATP Grant NNX08AH27G and funds provided by the University of Pennsylvania.


\section*{Appendix}

Recall from \eqref{13A} and \eqref{22C} that in $B=0$ gauge
\begin{equation}
ds^{2}={\cal{C}}(-\alpha z+{\cal{D}})^{1/3}dx^
{\mu}dx^{\nu}\eta_{\mu\nu}+dz^{2} 
\label{A1}
\end{equation}
where $z \in [0,\pi\rho]$.
Applying the formalism presented in  \cite{Goon:2011qf} to this heterotic background metric,
we find in terms of the ``brane-bending'' worldvolume scalar field $\pi(\sigma^{\mu})$ with $\mu=0,\dots,3$ that
\begin{eqnarray}
&&\sqrt{-h}|_{u=0}=f^4\sqrt{1+{1\over f^2}(\partial\pi)^2},\label{A2} \\
 &&{\hat{K}} = \gamma \Big(f^{-1}f'(5-\gamma^2)-f^{-2}[\Pi]+f^{-4}\gamma^2[\pi^3] \Big),\label{A3}\\
&&{\hat{R}}^{(4)} = \gamma f^{-4}\Big(\gamma \Big[[\Pi]^2-[\Pi^2] +2{\gamma^2\over f^2}\left(-[\Pi][\pi^3]+[\pi^4]\right)\Big] \label{A4}\\
&&\qquad +6{f^3f''\over \gamma}\big(-1+\gamma^2\big)
+2\gamma ff'\Big[-4[\Pi]+{\gamma^2\over f^2}\big(f^2[\Pi]+4[\pi^3]\big)\Big]  \nonumber \\
&&\qquad-6{(ff')^2\over \gamma}\big(1-2\gamma^2+\gamma^4\big)  \Big) \nonumber
\end{eqnarray}
where $\partial=\frac{\partial}{\partial\sigma^{\mu}}$, contraction is with respect to a flat metric $\eta_{\mu\nu}$,
\begin{eqnarray}
&&f= \mathcal{C}^{1/2}\left[-\alpha (z_0+\pi) + \mathcal{D}\right]^{1/6} \ , \label{A5}\\
&& \gamma = \frac{1}{\sqrt{1 + (\partial\pi)^2/f^2}} \ , \label{A6}
\end{eqnarray}
$z_{0}$ is the location of the kink hypersurface when $\epsilon \rightarrow 0$ and $'=\frac{\partial}{\partial \pi}$. 
Here $\Pi = \partial_\mu \partial_\nu \pi$, square brackets denote tracing with respect to $\eta_{\mu\nu}$ and $[\pi^n] = \partial \pi \cdot \Pi^{n-2} \cdot \partial \pi$. For example, $[\pi^3] = \partial_\mu \pi \partial^\mu \partial^\nu \pi \partial_\nu \pi$. \\


\begin{thebibliography}{99}

\bibitem{Aganagic:1996nn} 
  M.~Aganagic, C.~Popescu and J.~H.~Schwarz,
  ``Gauge invariant and gauge fixed D-brane actions,''
  Nucl.\ Phys.\ B {\bf 495}, 99 (1997)
  [hep-th/9612080].

\bibitem{Adawi:1997sq} 
  T.~Adawi, M.~Cederwall, U.~Gran, M.~Holm and B.~E.~W.~Nilsson,
  ``Superembeddings, nonlinear supersymmetry and five-branes,''
  Int.\ J.\ Mod.\ Phys.\ A {\bf 13}, 4691 (1998)
  [hep-th/9711203].

\bibitem{Sorokin:1999jx} 
  D.~P.~Sorokin,
  ``Superbranes and superembeddings,''
  Phys.\ Rept.\  {\bf 329}, 1 (2000)
  [hep-th/9906142].
  

\bibitem{Howe:2000vk} 
  P.~S.~Howe, A.~Kaya, E.~Sezgin and P.~Sundell,
  ``Codimension one-branes,''
  Nucl.\ Phys.\ B {\bf 587}, 481 (2000)
  [hep-th/0001169].
  
  
\bibitem{Derendinger:2000gy} 
  J.~-P.~Derendinger and R.~Sauser,
  ``A Five-brane modulus in the effective N=1 supergravity of M theory,''
  Nucl.\ Phys.\ B {\bf 598}, 87 (2001)
  [hep-th/0009054].
  
  
\bibitem{Howe:2001wc} 
  P.~S.~Howe and U.~Lindstrom,
  ``Kappa symmetric higher derivative terms in brane actions,''
  Class.\ Quant.\ Grav.\  {\bf 19}, 2813 (2002)
  [hep-th/0111036].
  
  
\bibitem{Cheung:2004sa} 
  Y.~-K.~E.~Cheung, M.~Laidlaw and K.~Savvidy,
  ``Open string gravity?'',
  JHEP {\bf 0412}, 028 (2004)
  [hep-th/0406245].
  
  
\bibitem{Belyaev:2010as} 
  D.~V.~Belyaev and T.~G.~Pugh,
  ``The Supermultiplet of boundary conditions in supergravity,''
  JHEP {\bf 1010}, 031 (2010)
  [arXiv:1008.1574 [hep-th]].
  
  
\bibitem{George:2009jn} 
  D.~P.~George and R.~R.~Volkas,
  ``Dynamics of the infinitely-thin kink,''
  Phys.\ Lett.\ B {\bf 704}, 646 (2011)
  [arXiv:0911.0538 [hep-th]].


\bibitem{Gregory:1990pm} 
  R.~Gregory,
  ``Effective actions for bosonic topological defects,''
  Phys.\ Rev.\ D {\bf 43}, 520 (1991).

\bibitem{Carter:1994ag} 
  B.~Carter and R.~Gregory,
  ``Curvature corrections to dynamics of domain walls,''
  Phys.\ Rev.\ D {\bf 51}, 5839 (1995)
  [hep-th/9410095].
  
\bibitem{KOS}   
 J.~Khoury, B.A.~Ovrut and J.~Stokes,
 ``The Worldvolume Action of Kink Solitons in AdS Spacetime,''
 arXiv:1203.4562v1 [hep-th] (2012).
 
\bibitem{Dvali:2000hr} 
  G.~R.~Dvali, G.~Gabadadze and M.~Porrati,
  ``4-D gravity on a brane in 5-D Minkowski space,''
  Phys.\ Lett.\ B {\bf 485}, 208 (2000)
  [hep-th/0005016].

 
   
\bibitem{Nicolis:2008in} 
  A.~Nicolis, R.~Rattazzi and E.~Trincherini,
  ``The Galileon as a local modification of gravity,''
  Phys.\ Rev.\ D {\bf 79}, 064036 (2009)
  [arXiv:0811.2197 [hep-th]].
  
  \bibitem{Trodden:2011xh} 
  M.~Trodden and K.~Hinterbichler,
  ``Generalizing Galileons,''
  Class.\ Quant.\ Grav.\  {\bf 28}, 204003 (2011)
  [arXiv:1104.2088 [hep-th]].
  
    
\bibitem{Khoury:2011da} 
  J.~Khoury, J.~-L.~Lehners and B.~A.~Ovrut,
  ``Supersymmetric Galileons,''
  Phys.\ Rev.\ D {\bf 84}, 043521 (2011)
  [arXiv:1103.0003 [hep-th]].

\bibitem{Endlich:2010zj} 
  S.~Endlich, K.~Hinterbichler, L.~Hui, A.~Nicolis and J.~Wang,
  ``Derrick's theorem beyond a potential,''
  JHEP {\bf 1105}, 073 (2011)
  [arXiv:1002.4873 [hep-th]].
  
\bibitem{deRham:2010eu} 
  C.~de Rham and A.~J.~Tolley,
  ``DBI and the Galileon reunited,''
  JCAP {\bf 1005}, 015 (2010)
  [arXiv:1003.5917 [hep-th]].
  
\bibitem{Hinterbichler:2010xn} 
  K.~Hinterbichler, M.~Trodden and D.~Wesley,
  ``Multi-field galileons and higher co-dimension branes,''
  Phys.\ Rev.\ D {\bf 82}, 124018 (2010)
  [arXiv:1008.1305 [hep-th]].
  
\bibitem{Goon:2010xh} 
  G.~L.~Goon, K.~Hinterbichler and M.~Trodden,
  ``Stability and superluminality of spherical DBI galileon solutions,''
  Phys.\ Rev.\ D {\bf 83}, 085015 (2011)
  [arXiv:1008.4580 [hep-th]].
  
  \bibitem{Duff:1994an} 
  M.~J.~Duff, R.~R.~Khuri and J.~X.~Lu,
  ``String solitons,''
  Phys.\ Rept.\  {\bf 259}, 213 (1995)
  [hep-th/9412184].
  
\bibitem{Stelle:1996tz} 
  K.~S.~Stelle,
  ``Lectures on supergravity p-branes,''
  In *Trieste 1996, High energy physics and cosmology* 287-339
  [hep-th/9701088].

\bibitem{a} 
P.~Horava, E.~Witten, 
``Eleven-Dimensional Supergravity on a Manifold with Boundary'',
Nucl.\ Phys.\ B{\bf 475} :94-114 (1996), 
[hep-th/9603142].

\bibitem{b} 
A.~Lukas, B.~A.~Ovrut, K.~S.~Stelle, D.~Waldram,
``The Universe as a Domain Wall'',
 Phys. \ Rev. \ D{\bf59} :086001(1999),
[hep-th/9803235].

\bibitem{c} 
A.~Lukas, B.~A.~Ovrut, K.~S.~Stelle and D.~Waldram,
``Heterotic M-Theory in Five Dimensions'',
Nucl.\ Phys.\  B {\bf 552}, 246 (1999),
[arXiv:hep-th/9806051].

\bibitem{d} 
A.~Lukas, B.~A.~Ovrut, D.~Waldram,
``Non-Standard Embedding and Five-Branes in Heterotic M-Theory'',
Phys. \ Rev.\ D{\bf 59} (1999) 106005,
arXiv:hep-th/9808101[hep-th].

\bibitem{e} 
  A.~Lukas, B.~A.~Ovrut and D.~Waldram,
  ``Five-branes and Supersymmetry Breaking in M-Theory,''
  JHEP {\bf 9904}, 009 (1999)
  [hep-th/9901017].
  
\bibitem{f} 
  R.~Donagi, B.~A.~Ovrut and D.~Waldram,
  ``Moduli spaces of five-branes on elliptic Calabi-Yau threefolds,''
  JHEP {\bf 9911}, 030 (1999)
  [hep-th/9904054].
  
\bibitem{Donagi:1998xe} 
  R.~Donagi, A.~Lukas, B.~A.~Ovrut and D.~Waldram,
  ``Nonperturbative vacua and particle physics in M theory,''
  JHEP {\bf 9905}, 018 (1999)
  [hep-th/9811168].
  
\bibitem{Donagi:1999gc} 
  R.~Donagi, A.~Lukas, B.~A.~Ovrut and D.~Waldram,
  ``Holomorphic vector bundles and nonperturbative vacua in M theory,''
  JHEP {\bf 9906}, 034 (1999)
  [hep-th/9901009].
  
\bibitem{Buchbinder:2002ji} 
  E.~Buchbinder, R.~Donagi and B.~A.~Ovrut,
  ``Vector Bundle Moduli and Small Instanton Transitions,''
  JHEP {\bf 0206}, 054 (2002)
  [hep-th/0202084].
  
\bibitem{Buchbinder:2002pr} 
  E.~I.~Buchbinder, R.~Donagi and B.~A.~Ovrut,
  ``Vector bundle moduli superpotentials in heterotic superstrings and M theory,''
  JHEP {\bf 0207}, 066 (2002)
  [hep-th/0206203].

\bibitem{A} 
  R.~Donagi, B.~A.~Ovrut, T.~Pantev and D.~Waldram,
  ``Standard Models From Heterotic M Theory,''
  Adv.\ Theor.\ Math.\ Phys.\  {\bf 5}, 93 (2002)
  [hep-th/9912208].
 
\bibitem{B} 
  R.~Donagi, B.~A.~Ovrut, T.~Pantev and D.~Waldram,
  ``Standard Model Bundles on Nonsimply Connected Calabi-Yau Threefolds,''
  JHEP {\bf 0108}, 053 (2001)
  [hep-th/0008008].
  
\bibitem{C} 
  R.~Donagi, B.~A.~Ovrut, T.~Pantev and D.~Waldram,
  ``Standard Model Bundles,''
  Adv.\ Theor.\ Math.\ Phys.\  {\bf 5}, 563 (2002)
  [math/0008010 [math-ag]].
  
\bibitem{D} 
  R.~Donagi, B.~A.~Ovrut, T.~Pantev and R.~Reinbacher,
  ``SU(4) instantons on Calabi-Yau Threefolds with Z(2) x Z(2) Fundamental Group,''
  JHEP {\bf 0401}, 022 (2004)
  [hep-th/0307273].
 
   
 \bibitem{g}   
 V.~Braun, Y.~H.~He, B.~A.~Ovrut, and T.~Pantev,
 ``The Exact MSSM Spectrum from String Theory'', 
JHEP0605:043,2006, 
[arXiv:hep-th/0512177].

\bibitem{h} 
L.~B.~Anderson, J.~Gray, Y.~H.~He and A.~Lukas,  
 ``Exploring Positive Monad Bundles And A New Heterotic Standard Model'',  
arXiv:0911.1569 [hep-th].
  
\bibitem{Braun:2005zv} 
  V.~Braun, Y.~-H.~He, B.~A.~Ovrut and T.~Pantev,
  ``Vector Bundle Extensions, Sheaf Cohomology, and the Heterotic Standard Model,''
  Adv.\ Theor.\ Math.\ Phys.\  {\bf 10}, 4 (2006)
  [hep-th/0505041].


\bibitem{Antunes:2002hn} 
  N.~D.~Antunes, E.~J.~Copeland, M.~Hindmarsh and A.~Lukas,
  ``Kinky brane worlds,''
  Phys.\ Rev.\ D {\bf 68}, 066005 (2003)
  [hep-th/0208219].

\bibitem{Goon:2011qf} 
  G.~Goon, K.~Hinterbichler and M.~Trodden,
  ``Symmetries for Galileons and DBI scalars on curved space,''
  JCAP {\bf 1107}, 017 (2011)
  [arXiv:1103.5745 [hep-th]].


\end{thebibliography}
\end{document}